\documentclass[twocolumn,superscriptaddress,floatfix]{revtex4-2}
\usepackage{gnuplottex}
\usepackage{mathtools}
\usepackage{float}
\usepackage{array}
\usepackage[english]{babel}
\usepackage[utf8]{inputenc}
\usepackage{amsmath}
\usepackage[table]{xcolor}
\usepackage{amsfonts}
\usepackage{graphicx}
\usepackage{dsfont}
\usepackage[colorlinks=true, linkcolor=blue, citecolor=dred]{hyperref}
\usepackage{algorithm}
\usepackage[noend]{algpseudocode}
\usepackage{bbold}
\usepackage{mathtools}
\usepackage{amssymb}
\usepackage{bm}
\usepackage{bbm}
\usepackage{tabularx, booktabs}
\usepackage{xspace}
\usepackage{listings}
\usepackage{tikz}
\usepackage{bbding}
\usepackage{physics}
\usepackage{amsmath}
\usepackage{soul}
\usepackage{caption}
\usetikzlibrary{calc, positioning, quantikz}
\usepackage[noend]{algpseudocode}
\newcolumntype{Y}{>{\centering\arraybackslash}X}
\definecolor{dgreen}{rgb}{0,.5,0}
\definecolor{dblue}{rgb}{0,0,.5}
\definecolor{dred}{rgb}{0.5,0,.5}

\newcommand{\bmq}{\bm{q}}
\newcommand{\psh}[2]{\ensuremath{\langle #1|#2\rangle}\xspace}

\newcommand{\bmkappa}{{\bm \kappa}}
\newcommand{\bmtheta}{{\bm \theta}}

\newcommand{\rmA}{{\rm A}}
\newcommand{\rmB}{{\rm B}}

\DeclareUnicodeCharacter{2212}{-}
\DeclareUnicodeCharacter{2009}{\,}

\DeclareRobustCommand{\martinfix}[2]{\textcolor{green}{\st{#1} #2}}

\begin{document}

\title{Transformation-free generation of a quasi-diabatic representation from the state-average orbital-optimized variational quantum eigensolver}

\author{Silvie Illésová}
\affiliation{IT4Innovations, VSB – Technical University of Ostrava, 17. listopadu 2172/15, 708 00 Ostrava-Poruba, Czech Republic}
\author{Martin Beseda}
\affiliation{Dipartimento di Ingegneria e Scienze dell'Informazione e Matematica, Universit\`a dell'Aquila, via Vetoio,
I-67010 Coppito-L'Aquila, Italy.}
\author{Saad Yalouz}
\affiliation{Laboratoire de Chimie Quantique, Institut de Chimie,
CNRS/Université de Strasbourg, 4 rue Blaise Pascal, 67000 Strasbourg, France}
\author{Benjamin Lasorne}
\thanks{corresponding author: \href{mailto:benjamin.lasorne@umontpellier.fr}{benjamin.lasorne@umontpellier.fr}}
\affiliation{ICGM, Univ Montpellier, CNRS, ENSCM, Montpellier, France}
\author{Bruno Senjean}
\thanks{corresponding author: \href{mailto:bruno.senjean@umontpellier.fr}{bruno.senjean@umontpellier.fr}}
\affiliation{ICGM, Univ Montpellier, CNRS, ENSCM, Montpellier, France}

\date{\today}

\begin{abstract}
In the present work, we examine how the recent quantum-computing algorithm known as the state-average orbital-optimized variational quantum eigensolver (SA-OO-VQE), viewed within the context of quantum chemistry as a type of multiconfiguration self-consistent field (MCSCF) electronic-structure approach, exhibits a propensity to produce an \emph{ab initio} quasi-diabatic representation ``for free'' if considered as a least-transformed block-diagonalization procedure, as alluded to in our previous work [S. Yalouz \emph{et al.},
J. Chem. Theory Comput. 18 (2022) 776] and thoroughly assessed herein. 
To this end, we introduce intrinsic and residual descriptors of diabaticity and re-explore the definition and linear-algebra properties -- as well as their consequences on the vibronic nonadiabatic couplings -- of an optimal diabatic representation within this context, and how much one may deviate from it.
Such considerations are illustrated numerically on the prototypical case of formaldimine, which presents a well-known conical intersection between its ground and first-excited singlet electronic
states. 
\end{abstract}

\maketitle


\section{Introduction}\label{sec:intro}

Ultrafast (subpicosecond) photochemical and photophysical `energy/charge/matter/information'-transfer processes induced upon light absorption by molecules within the UV-visible domain are ubiquitous. 
Subsequent radiationless processes, in particular, are governed by so-called vibronic nonadiabatic couplings (NACs) among several adiabatic electronic states. 
They are mediated by molecular vibrations and how these affect the electronic system when the nuclei move from their equilibrium positions. 
In other words, vibronic NACs -- when they are large -- imply a theoretical description that goes beyond the Born-Oppenheimer approximation (BOA) \cite{bae06,dom04,dom11,las11:460}. 
Their effects have been observed and rationalized, both for time-resolved experiments (femtochemistry; see, \emph{e.g.}, Ref. \cite{zew96:12701}) and energy-resolved ones (photoabsorption or photoelectron spectroscopy; see, \emph{e.g.}, Refs. \cite{raa99:936, cat01:2088}).

Any faithful theoretical/computational description (modelling and simulation) of such processes beyond BOA requires -- in principle -- a quantum treatment of the nuclear motion (molecular quantum dynamics) \cite{gat17,gon21} because the vibrational system experiences a dynamical regime that is ``as quantum'' as that of the electronic system as regards energy and time scales. 
This has been one of the main incentives for the rise of nuclear wavepacket methods within the field of theoretical chemistry over the past few decades, incarnated in particular with the now golden standard known as the `multiconfiguration time-dependent Hartree' (MCTDH) method \cite{bec00:1,wor20:107040}. 

Meanwhile, alternative approximate treatments relying on statistical ensembles of nuclear trajectories attached to quantum superpositions of electronic states, rather than molecular wavepackets -- known as mixed quantum-classical and sometimes semiclassical -- have been devised over the years. 
The most popular one is perhaps the Tully `trajectory surface hopping' (TSH) approach (see, \emph{e.g.}, Ref. \cite{mer23:1827} for a recent review comparing various implementations of the NACs). 
In practice, both types of approaches exhibit very different interfacial connections toward electronic-structure computations.

While grid-based wavepacket methods rely on the availability, hence the preconstruction, of a global functional representation of the electronic Hamiltonian matrix at any molecular geometry (also known as a `fit', much as in a parameterized force field), trajectory-based methods -- known as `nonadiabatic molecular dynamics' (NAMD) \cite{tav15:792} -- can benefit from direct, \emph{ab initio}, local evaluations of energy derivatives and NACs on-the-fly, within a spirit similar to `Born-Oppenheimer molecular dynamics' (BOMD) or `\emph{ab initio} molecular dynamics' (AIMD) \cite{mar09}. 
A rich variety of intermediate approaches, aiming at combining the pros and cons of the two fully-quantum and mixed quantum-classical extremes, often Gaussian-based, have been proposed over the past decades. 
This has become a very active field of research, with the common aim of proposing a rational hierarchy of methods \cite{las11:460,cre18:7026,ago19:e1417,gom24:1829}.
A full exposition of them is beyond the scope of the present work. 

Now, in practice, it must be understood that wavepacket methods -- either grid-based or Gaussian-based -- are quantum, hence not pointwise with respect to the nuclear coordinates. 
They thus involve eventually the evaluation of multidimensional Hamiltonian matrix elements. 
As such, any singular or ill-behaved operator is not to be welcome, being a potential source for numerical difficulties upon local differentiation and/or integration over a spatial distribution. 
This is the very reason why so-called (quasi)diabatic representations have become preferred in this context.
The problem is well-known and concerns in particular the treatment of tamed vibronic NACs. 
This aspect is one of the main focuses of the present work.

In a nutshell, the most evident representation of the electronic problem is the adiabatic one (eigensolutions of the electronic Hamiltonian at any molecular geometry). 
This is the very purpose of quantum chemistry. 
However, when two well-defined adiabatic eigenstates become degenerate at a certain geometry, known as a conical intersection, which is a common situation of physical interest for radiationless decay, they exhibit a divergent NAC between them and the problem becomes mathematically ill-defined. 
A solution to this consists of invoking a global geometry-dependent unitary change of basis (known as the `adiabatic-to-diabatic transformation' -- ADT), which is meant to generate a well-behaved electronic basis set with vanishing NACs at all molecular geometries.
Such an ADT is not unique and rarely exists globally in a strict sense. This has been the subject of some abundant literature in the field during the 1980s (see, \emph{e.g.}, Ref. \cite{bae06} and references therein for a quite exhaustive compilation). 

A plethora of practical approximate solutions has been proposed over the years, typically known as `quasi-diabatizations'. 
A complete overview is beyond the scope of the present work. 
Let us mention that there are essentially three types of approaches: (i) NAC-based, with line-integration of the vector criterion of ``diabaticity'' (see, \emph{e.g.}, Ref. \cite{ric15:12457} for a recent implementation in the context of Gaussian-based direct quantum dynamics); 
(ii) property-based, with smooth and minimal variations of expectation values of operators (observables), which somewhat includes the famous ``diabatization by ansatz'' where the electronic Hamiltonian matrix elements are sought from the onset as regular functions (low-degree polynomials) of the nuclear coordinates  (see, \emph{e.g.}, Refs. \cite{raa99:936,cat01:2088} for two milestone illustrations and \cite{vie21:084302} for a recent extension making use of neural-network techniques); 
(iii) wavefunction-based, according to a maximal overlap criterion of the electronic states at various molecular geometries (see, \emph{e.g.}, Refs. \cite{zha21:1885,shu22:992} for recent considerations about the Householder compression with valence-bond wavefunctions and some related approaches for the automatization of the global phase-consistency issue).

In the present work, we shall investigate an approach of the third kind. 
It strongly relies on the concept of a least-transformed block-diagonalization, which was first proposed as an avenue by Cederbaum \emph{et al.} in Refs. \cite{pac88:7367,ced89:2427} and further explored in practice by Werner \emph{et al.} in Refs. \cite{wer88:3139,sim99:4523}.
The formal basis invokes a constrained singular-value decomposition (SVD) between the model and target subspaces.
Its connection with effective Hamiltonian theory and quasidegenerate perturbation theory (QDPT) \cite{des60:321,kle74:786,bran79:207} is evident and gave rise to strong activity in the field at that time (see, \emph{e.g.}, Refs. \cite{spi84:1259,cim85:3073,gad86:4872,des87:1429}). 
Interestingly enough, some recent works along this line have rejuvenated this least-transformation concept through linear-algebra lenses (see, \emph{e.g.}, Refs. \cite{ric20:154108,zha21:1885}, upon realizing that this is a unitary Procrustes problem and/or a Householder compression problem.

All wavefunction-based approaches cited above can be viewed as utilizing some post-processing of the \emph{ab initio} adiabatic results with a back-transformation achieving the constraint that the ADT target states overlap maximally with the model states. 
While elegant, and even obvious, in the context of `configuration interaction' (CI) methods -- including all kinds of `multiconfiguration self-consistent field' (MCSCF) variants with `orbital optimization' (OO), such approaches have always suffered from technical difficulties, essentially related to phase inconsistency stemming from gauge freedom. 
It turns out that diabatizing the many-body CI problem can be addressed easily for some given set of orbitals, but the hidden difficulty is then transferred to the one-body problem, for which the orbitals have to remain ``as diabatic as possible'' from a given molecular geometry to another. 

Yet, the CI problem is the first source of trouble, and it is the primary focus of the present work. 
We shall show that `quantum-computing' (QC) algorithms provide by design a totally new perspective of foremost interest here. 
The most important point is the following. 
From a classical-computing perspective, the CI many-body variational problem is merely a linear eigenproblem (diagonalization), which is to be solved fully (Jacobi) or partly (Lanczos or Davidson). 
In the latter case, the target subspace (the ``ensemble'') must overlap strongly with the model subspace (the ``guess''). 
A state-specific (SS) or a state-average (SA) MCSCF approach will eventually do the same at the end of a nonlinear self-consistent field (SCF) -- coupled: CI+OO -- iterative procedure where the SA-ensemble or SS-state are expected to match the target for a partial diagonalization. 
In all situations, a full or partial diagonalization algorithm is involved so as to produce a subspace of adiabatic solutions. 
Now, the corresponding QC `variational quantum eigensolver' (VQE) algorithms work in a different way when an ensemble is involved. 

Ensemble VQE algorithms have been originally proposed
by Nakanishi {\it et al.}~\cite{nakanishi2019subspace}. 
When all states of the ensemble are equal-weighted, we call it SA-VQE.
The SA-VQE algorithm has no standard classical-computing counterpart (this would be called SA-CI with given orbitals): its variational objective is only to minimize an ensemble energy, according to the Theophilou and Gross--Oliveira--Kohn principle, which is the generalization of the Rayleigh--Ritz variational principle to an ensemble of states~\cite{theophilou1979energy,gross1988rayleigh,ding2024ground}.

The SA-OO-VQE algorithm is similar to SA-VQE, but with OO, and thus simulates SA-MCSCF~\cite{yal21:024004,yal22:776}. 
Either way, their variational objectives are not to provide adiabatic states (full or partial diagonalization), but only an optimal eigensubspace (block-diagonalization).
This is reminiscent of the variational interpretation of Hartree--Fock \cite{sza96}, where the actual objective is not canonical \emph{per se}, but canonicalization occurs to be an efficient procedure. 
Hence, partial diagonalization is a stronger condition than merely fulfilling the ensemble variational principle and could thus be termed post-variational.

In other words, both SA-VQE and SA-OO-VQE with equal weights within an ensemble are meant to minimize the trace of a target matrix block and no more. 
As such, they are supposed to block-diagonalize the many-body Hamiltonian matrix in the basis of configuration-state functions (CSFs), or Slater determinants (SDs) if no spin-restriction is considered. 
Note that considering ordered weights instead of equal weights allows extracting, in principle, the eigenstates directly~\cite{gross1988rayleigh,nakanishi2019subspace}.
Practical implementations based on a variational unitary coupled-cluster (UCC) ansatz -- and in particular the generalized UCC with singles and doubles (GUCCSD) one, used in the present work, also referred to as a state-agnostic quantum
UCC ansatz~\cite{mcclean2016theory} -- can be parametrized in such a way that they achieve the true variational objective (block-diagonalization) only, or it together with the post-variational objective (partial diagonalization), with extra parameters.

Because both steps can be disconnected, we now have a way to achieve some kind of \emph{ab initio} and \emph{a priori} diabatization that is transformation-free and does not require any \emph{a posteriori} post-processing ADT. 
While there is no definitive proof that the optimization algorithm is a least-transformation minimization, such a promising property has already been noticed heuristically by us in previous works \cite{yal21:024004,yal22:776}. 
Its more detailed investigation is the subject of the present work and is presented as a numerical proof of principle in the case of the formaldimine molecule.
To this end, various types of ``diabaticity'' descriptors are introduced and analyzed. 

The paper outline is organized as follows.
In Sec.~\ref{sec:theory} we provide details on the essential theoretical background;
the computational method is exposed in Sec.~\ref{sec:compdet};
results and discussion are shown in Sec.~\ref{sec:resdisc};
conclusions and outlook are to be found in Sec.~\ref{sec:conclu}; 
we terminate with a substantial list of appendices meant to facilitate and expand the connections among the various concepts and relations involved in the main text.

\section{Conceptual background}\label{sec:theory}

The goal of this section is to provide the necessary and sufficient notations and formal details for a clear understanding of the non-trivial concept of (quasi-)diabatic states and of their relation to adiabatic ones within an MCSCF context.
We aim to guide the reader through these fundamental notions, focusing specifically on the case of a two-state problem, which is the focus of the present work.
Given the vastness of the literature on the subject, we restrict our discussion to the essential aspects, which are summarized in Fig.~\ref{fig:summary} serving as a graphical companion, while additional details are provided in the Appendices for those interested.

\begin{figure*}
    \centering
    \resizebox{\textwidth}{!}{
    \includegraphics{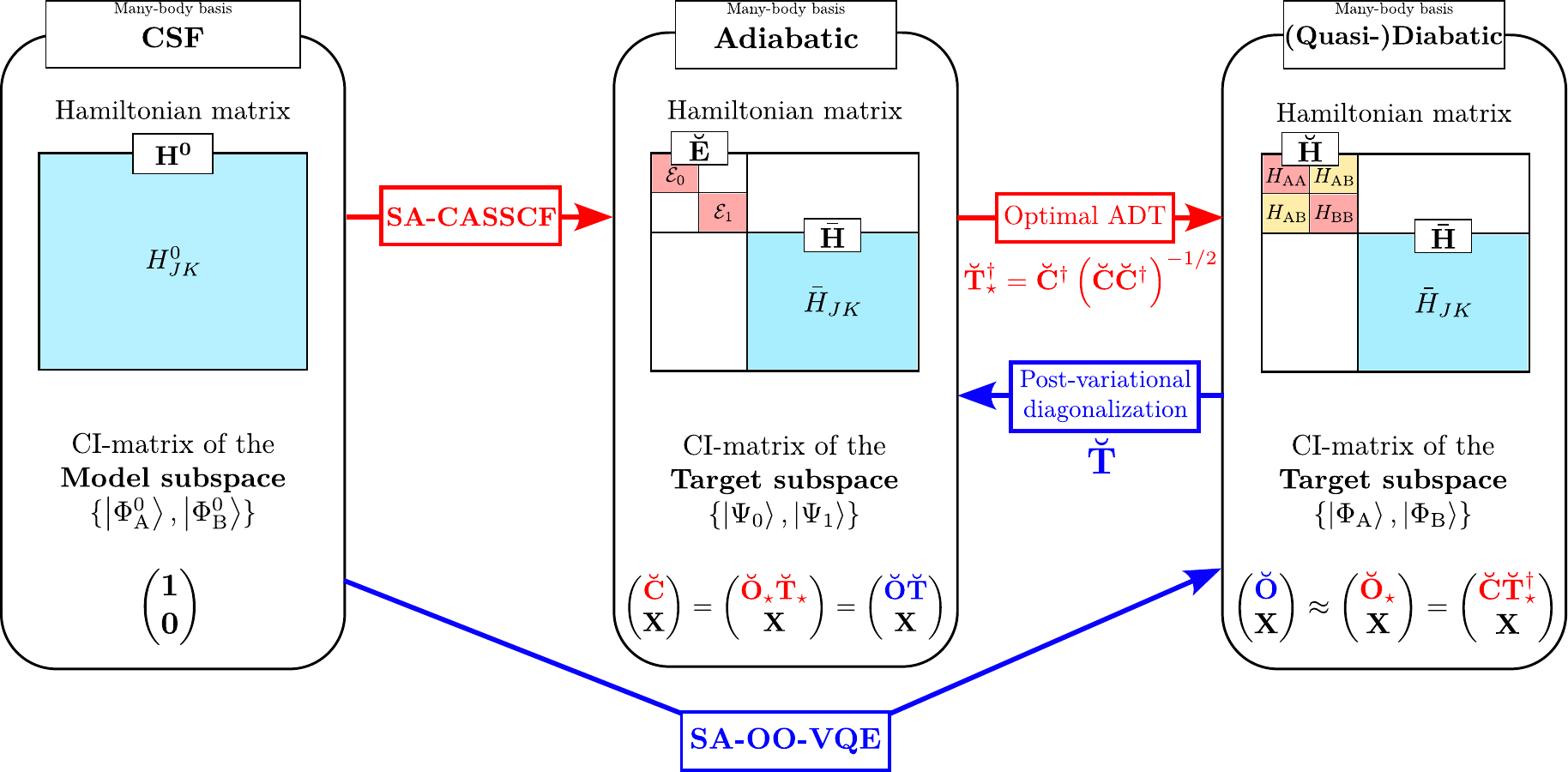}
    }
    \caption{Schematic graphical summary of the essential quantities involved in this work and their relations. In red, the usual path to get (quasi-)diabatic states from SA-CASSCF. In blue, the path taken in this work using SA-OO-VQE. `CI-matrix' refers to the first two CI-coefficient column-vectors, corresponding to the two states of the model/target subspaces.}
    \label{fig:summary}
\end{figure*}

\subsection{Nonadiabatic couplings within a two-state target subspace}\label{subsec:NAC2states}

Let us denote $\bm{q}$ the set of internal coordinates that determine the `molecular geometry'. 
For each value of $\bm{q}$, we have a standard quantum-chemical electronic Hamiltonian 
(including nuclear-nuclear potential-energy repulsion, but discarding the nuclear kinetic-energy term), denoted $\hat{H}^{\rm{el}}[\bm{q}]$, and experiencing a smooth parametric dependence with respect to $\bm{q}$-variations.

Herein, we shall restrict ourselves 
to the first two singlet many-body electronic eigenstates of this operator (assuming an even number of electrons), which define a target subspace within the spirit of a two-state effective-Hamiltonian description. 
For the moment, let us consider that we can solve the Schr{\"o}dinger eigenproblem exactly. 
The first two -- so-called adiabatic -- singlet states (also labeled $\rm S_0$ and $\rm S_1$) are the supposedly-exact 
eigenstates of $\hat{H}^{\rm{el}}[\bm{q}]$ at any $\bm{q}$, 
\begin{subequations}
\label{eq:idealtise}
\begin{align}
&\forall \bm{q},\quad \hat{H}^{\rm{el}}[\bm{q}] \ket{\Psi_0;\bm{q}} = \mathcal{E}_0(\bm{q}) \ket{\Psi_0;\bm{q}} \quad, \\
&\forall \bm{q},\quad \hat{H}^{\rm{el}}[\bm{q}] \ket{\Psi_1;\bm{q}} = \mathcal{E}_1(\bm{q}) \ket{\Psi_1;\bm{q}} \quad,
\end{align}
\end{subequations}
where we assume adiabatic ordering: $\forall \bm{q}$, $\mathcal{E}_0(\bm{q}) \leq \mathcal{E}_1(\bm{q})$. 

They both span the `target Hilbert subspace'. 
The semicolon notation indicates their 
parametric dependence with respect to $\bm{q}$ (expected to be smooth, except at -- and somewhere around -- conical intersections). 
Dirac's `ket' notation implicitly defines the electronic degrees of freedom regarding Hilbert integration. 
From now on, let us drop the `$\forall \bm{q}$' specification and make it implicit.

The electronic Hamiltonian is Hermitian so that the two eigenstates are orthogonal and they can 
be chosen to be normalized. 
The previous partial eigenproblem can thus be recast in matrix form as, $\forall \alpha,\beta\in\{0,1\}$,
\begin{subequations}
\label{eq:idealeigen}
\begin{align}
& [\bold{E}(\bm{q})]_{\alpha\beta}=\mel{\Psi_\alpha;\bm{q}}{\hat{H}^{\rm{el}}[\bm{q}]}{\Psi_\beta;\bm{q}}=E_{\alpha\beta}(\bm{q})=\mathcal{E}_\alpha(\bm{q})\delta_{\alpha\beta} \quad, \\
& [\bold{1}]_{\alpha\beta}=\braket{\Psi_\alpha;\bm{q}}{\Psi_\beta;\bm{q}}=\delta_{\alpha\beta} \quad,
\end{align}
\end{subequations}
where $\delta_{\alpha\beta}$ is the Kronecker symbol. 

Now, let us equip the many-body Hilbert space with a discrete but complete orthonormal basis set, $\{\ket{\Phi^0_J;\bm{q}}\}_{J=0}^\infty$.
It is assumed to vary as smoothly as possible with respect to $\bm{q}$, and thus be ``as diabatic as possible'' (this concept will be clarified later on). 
In practice, it will be made of a finite set of SDs -- or, better, singlet CSFs -- where the smooth $\bm{q}$-dependence only comes from the underlying 
molecular orbitals (via the `linear combination of atomic orbitals' (LCAO) coefficients and the `atomic orbital' (AO) overlaps; see Appendix~\ref{app:diaborbs}) and defines what is known as the one-body-nonremovable part of the NACs, also known as the CSF-NACs (see Appendix~\ref{app:NAC}).

Before going further, we must introduce a fundamental predicament as regards the present work. 
The first two many-body basic 
states, denoted $\ket{\Phi^0_0;\bm{q}}\equiv\ket{\Phi^0_\rmA;\bm{q}}$ and $\ket{\Phi^0_1;\bm{q}}\equiv\ket{\Phi^0_\rmB;\bm{q}}$, will be assumed from the onset to span what is known as the `model Hilbert subspace'. 
It has to be a ``good guess'' of the `target Hilbert subspace', based on educated chemical intuition (choice of the dominant CSFs). 
This may seem abstract for the moment but we shall define actual descriptors for this further on.

Let us put the OO problem on hold for a moment
and assume numerical 
completeness with respect to the CI problem of a finite CSF basis set $\{\ket{\Phi^0_J;\bm{q}}\}_{J=0}^{N-1}$ (with $N>0$). 
The target adiabatic states can be unitarily expanded as
\begin{subequations}
\begin{align}
& \ket{\Psi_0;\bm{q}}=\sum_{J=0}^{N-1} \ket{\Phi^0_J;\bm{q}} C_{J0}(\bm{q}) \quad, \\
& \ket{\Psi_1;\bm{q}}=\sum_{J=0}^{N-1} \ket{\Phi^0_J;\bm{q}} C_{J1}(\bm{q}) \quad,
\end{align}
\end{subequations}
where
\begin{subequations}
\begin{align}
& C_{J0}(\bm{q})=\braket{\Phi^0_J;\bm{q}}{\Psi_0;\bm{q}} \quad, \\
& C_{J1}(\bm{q})=\braket{\Phi^0_J;\bm{q}}{\Psi_1;\bm{q}} \quad.
\end{align}
\end{subequations}
This specifies the first two columns of a CI-coefficient matrix,
\begin{equation}
\label{eq:cmatstruct}
\begin{pmatrix}
\bold{\breve{C}}(\bm{q}) \\
\bold{X}(\bm{q})
\end{pmatrix} \quad, \quad
\bold{\breve{C}}(\bm{q}) = 
\begin{pmatrix}
C_{\rmA0}(\bm{q}) & C_{\rmA1}(\bm{q}) \\
C_{\rmB0}(\bm{q}) & C_{\rmB1}(\bm{q})
\end{pmatrix} 
\quad,
\end{equation}
which may be fully augmented to a unitary $(N \times N)$-$\bold{C}(\bm{q})$-matrix upon some arbitrary orthonormal completion, consistent with a subspace partition meant to achieve block-diagonalization (see Appendix~\ref{app:lowdin}).
Note that, throughout this paper, the ``breve'' symbol, $\bold{\breve{...}}$, will be used to specify target or model submatrices: namely, here, a $(2 \times 2)$-block.

From the above, we can recast the partial eigenproblem in matrix form as
\begin{subequations}
\label{eq:citise}
\begin{align}
& \sum_{K=0}^{N-1} H^0_{JK}(\bm{q}) C_{K0}(\bm{q}) = \mathcal{E}_0(\bm{q}) C_{J0}(\bm{q}) \quad, \\
& \sum_{K=0}^{N-1} H^0_{JK}(\bm{q}) C_{K1}(\bm{q}) = \mathcal{E}_1(\bm{q}) C_{J1}(\bm{q}) \quad,
\end{align}
\end{subequations}
where
\begin{equation}
[\bold{H}^0(\bm{q})]_{JK}=\mel{\Phi^0_J;\bm{q}}{\hat{H}^{\rm{el}}[\bm{q}]}{\Phi^0_K;\bm{q}} = H^0_{JK}(\bm{q}) \quad.
\end{equation}
In other words, and with evident notations for orthonormal completion to the full Hilbert space, we get
\begin{subequations}
\begin{align}
& \bold{C}^\dag(\bm{q}) \bold{H}^0(\bm{q}) \bold{C}(\bm{q}) = \bold{E}(\bm{q}) \quad, \\
& \bold{C}^\dag(\bm{q}) \bold{C}(\bm{q}) = \bold{1} \quad,
\end{align}
\end{subequations}
where $\bold{E}(\bm{q})$  is a block-diagonal matrix, with a diagonal $(2 \times 2)$-block for the first two indices, denoted $\bold{\breve{E}}(\bm{q})$, such that 
\begin{equation}
\bold{E}(\bm{q}) = 
\begin{pmatrix}
\bold{\breve{E}}(\bm{q}) & \bm0 \\
\bm0 & \bold{\bar{H}}(\bm{q})
\end{pmatrix} \quad,
\end{equation}
where 
\begin{equation}
\label{eq:cieigen}
\bold{\breve{E}}(\bm{q}) = 
\begin{pmatrix}
\mathcal{E}_0(\bm{q}) & 0 \\
0 & \mathcal{E}_1(\bm{q})
\end{pmatrix} \quad.
\end{equation}
The complementary block, denoted $\bold{\bar{H}}(\bm{q})$, is undetermined but is irrelevant for the present purpose. 

It must be understood here that the two eigenvalues, $\mathcal{E}_0(\bm{q})$ and $\mathcal{E}_1(\bm{q})$, in Eqs.~(\ref{eq:citise}) and (\ref{eq:cieigen}) 
are only the best variational approximations to both targeted exact eigenvalues in Eqs.~(\ref{eq:idealtise}) and (\ref{eq:idealeigen}) when considering a 
finite $N$-sized approximation of the many-body Hilbert space. 
Now, we may invoke the existence of an intermediate basis set of the target subspace, 
$\{\ket{\Phi_0;\bm{q}},\ket{\Phi_1;\bm{q}}\}\equiv\{\ket{\Phi_\rmA;\bm{q}},\ket{\Phi_\rmB;\bm{q}}\}$ (its practical meaning as regards quasi-diabatization through its relation with the model subspace will be clarified soon).
Assuming real-valued many-body wavefunctions, the adiabatic states can always be obtained from it in practice \emph{via} a final rotation through the post-variational diagonalization angle, $\varphi(\bm{q})$,
\begin{subequations}
\label{eq:rotation}
\begin{align}
& \ket{\Psi_0;\bm{q}} = \cos{\varphi(\bm{q})} \ket{\Phi_\rmA;\bm{q}} + \sin{\varphi(\bm{q})} \ket{\Phi_\rmB;\bm{q}} \quad, \\
& \ket{\Psi_1;\bm{q}} = -\sin{\varphi(\bm{q})} \ket{\Phi_\rmA;\bm{q}} + \cos{\varphi(\bm{q})} \ket{\Phi_\rmB;\bm{q}} \quad.
\end{align}
\end{subequations}
The corresponding change-of-basis $(2 \times 2)$-matrix will be denoted
\begin{equation}
\bold{\breve{T}}(\bm{q}) = 
\begin{pmatrix}
\cos{\varphi(\bm{q})} & -\sin{\varphi(\bm{q})} \\
\sin{\varphi(\bm{q})} & \cos{\varphi(\bm{q})}
\end{pmatrix} \quad.
\end{equation}

The $(2 \times 2)$-matrix representation of the electronic Hamiltonian 
within this basis will be denoted, $\forall J,K\in\{0,1\}\equiv\{\rmA,\rmB\}$, 
\begin{equation}
[\bold{\breve{H}}(\bm{q})]_{JK}=\mel{\Phi_J;\bm{q}}{\hat{H}^{\rm{el}}[\bm{q}]}{\Phi_K;\bm{q}} = H_{JK}(\bm{q}) \quad.
\end{equation}
The completion of it to the full matrix $\bold{H}(\bm{q})$ is as undetermined as that of $\bold{E}(\bm{q})$; their complements can be viewed as being identical. 
In other words, both full matrices are block-diagonal and differ only by their respective first $(2 \times 2)$-blocks. 
The latter are related \emph{via}
\begin{subequations}
\begin{align}
& \bold{\breve{T}}^\dag(\bm{q}) \bold{\breve{H}}(\bm{q}) \bold{\breve{T}}(\bm{q}) = \bold{\breve{E}}(\bm{q}) \quad, \\
& \bold{\breve{T}}^\dag(\bm{q}) \bold{\breve{T}}(\bm{q}) = \bold{1} \quad.
\end{align}
\end{subequations}

Since we are dealing with a two-state problem, we have access to explicit eigensolutions, which can take benefit from a reformulation in terms of Pauli matrices: $\boldsymbol\sigma_0\equiv\bm1$ and
\begin{equation}
\boldsymbol\sigma_1 =
\begin{pmatrix}
0 & 1 \\
1 & 0
\end{pmatrix}
\quad
\boldsymbol\sigma_2 =
\begin{pmatrix}
0 & -i \\
i & 0
\end{pmatrix}
\quad
\boldsymbol\sigma_3 =
\begin{pmatrix}
1 & 0 \\
0 & -1
\end{pmatrix}
\quad.
\end{equation}
Assuming real-valued matrix elements, the change-of-basis $(2 \times 2)$-matrix (supposed to be the inverse of the ADT when the intermediate basis is diabatic) and the Hamiltonian submatrices can be expressed as (special-orthogonal: rotation)
\begin{equation}
\bold{\breve{T}}(\bm{q})=e^{-i\varphi(\bm{q})\boldsymbol\sigma_2} = \cos{\varphi(\bm{q})}\bm1 - i\sin{\varphi(\bm{q})}\boldsymbol\sigma_2 \quad,
\end{equation}
and (real-symmetric)
\begin{subequations}
\begin{align}
& \bold{\breve{E}}(\bm{q}) = \mathcal{S}(\bm{q}) \bm1 + 0 \boldsymbol\sigma_1 - \mathcal{D}(\bm{q}) \boldsymbol\sigma_3
\quad, \\
& \bold{\breve{H}}(\bm{q}) = S(\bm{q}) \bm1 + W(\bm{q}) \boldsymbol\sigma_1 - D(\bm{q}) \boldsymbol\sigma_3
\quad,
\end{align}
\end{subequations}
where we define 
\begin{subequations}
\begin{align}
& S(\bm{q})=\frac{H_{\rmA\rmA}(\bm{q})+H_{\rmB\rmB}(\bm{q})}{2} \quad, \\
& D(\bm{q})=\frac{H_{\rmB\rmB}(\bm{q})-H_{\rmA\rmA}(\bm{q})}{2} \quad, \\
& W(\bm{q})=H_{\rmA\rmB}(\bm{q})=H_{\rmB\rmA}(\bm{q}) \quad,
\end{align}
\end{subequations}
and 
\begin{subequations}
\begin{align}
& \mathcal{S}(\bm{q})=\frac{\mathcal{E}_{0}(\bm{q})+\mathcal{E}_{1}(\bm{q})}{2} \quad, \\
& \mathcal{D}(\bm{q})=\frac{\mathcal{E}_{1}(\bm{q})-\mathcal{E}_{0}(\bm{q})}{2}
\geq 0 \quad.
\end{align}
\end{subequations}
Hence, from
\begin{equation}
\bold{\breve{H}}(\bm{q}) = e^{-i\varphi(\bm{q})\boldsymbol\sigma_2} \bold{\breve{E}}(\bm{q}) e^{i\varphi(\bm{q})\boldsymbol\sigma_2} \quad,
\end{equation}
and owing to
\begin{subequations}
\label{eq:pauli-simi}
\begin{align}
& e^{-i\varphi\boldsymbol\sigma_2} \bm1 e^{i\varphi\boldsymbol\sigma_2} = \bm1
\quad, \\
& e^{-i\varphi\boldsymbol\sigma_2} \boldsymbol\sigma_1 e^{i\varphi\boldsymbol\sigma_2} = \cos{2\varphi}\boldsymbol\sigma_1-\sin{2\varphi}\boldsymbol\sigma_3
\quad, \\
& e^{-i\varphi\boldsymbol\sigma_2} \boldsymbol\sigma_2 e^{i\varphi\boldsymbol\sigma_2} = \boldsymbol\sigma_2
\quad, \\
& e^{-i\varphi\boldsymbol\sigma_2} \boldsymbol\sigma_3 e^{i\varphi\boldsymbol\sigma_2} = \sin{2\varphi}\boldsymbol\sigma_1+\cos{2\varphi}\boldsymbol\sigma_3
\quad,
\end{align}
\end{subequations}
we obtain that the diagonalization angle and the two invariants satisfy
\begin{subequations}
\label{eq:polar}
\begin{align}
& \cos{2\varphi(\bm{q})}=\frac{D(\bm{q})}{\sqrt{(D(\bm{q}))^2+(W(\bm{q}))^2}} \quad, \\
& \sin{2\varphi(\bm{q})}=-\frac{W(\bm{q})}{\sqrt{(D(\bm{q}))^2+(W(\bm{q}))^2}} \quad, \\
& \mathcal{S}(\bm{q})=S(\bm{q}) \quad, \\
& \mathcal{D}(\bm{q})=\sqrt{(D(\bm{q}))^2+(W(\bm{q}))^2} \quad.
\end{align}
\end{subequations}
These have a similar structure to a Cartesian-to-polar change of coordinates from $(D,W)$ to $(\mathcal{D},-2\varphi)$.
As such, we get the local gradients (with respect to $\bm{q}$, denoted with $\bm\nabla$) as
\begin{subequations}
\label{eq:grads-nacs}
\begin{align}
& \bm\nabla \mathcal{D}(\bm{q}) = \frac{D(\bm{q}) \bm\nabla D(\bm{q}) + W(\bm{q}) \bm\nabla W(\bm{q})}{\sqrt{(D(\bm{q}))^2+(W(\bm{q}))^2}} \quad, \\
& -2\mathcal{D}(\bm{q})\bm\nabla \varphi(\bm{q}) = \frac{-W(\bm{q}) \bm\nabla D(\bm{q}) + D(\bm{q}) \bm\nabla W(\bm{q})}{\sqrt{(D(\bm{q}))^2+(W(\bm{q}))^2}} \quad.
\end{align}
\end{subequations}
Note that such local derivatives at a conical intersection, where $\mathcal{D}(\bm{q})=0$, are singular but can be replaced by directional derivatives.

Interestingly enough, this provides an explicit analytic expression for $\bm\nabla \varphi(\bm{q})$ and thus an exact way for evaluating it, provided that we can compute the values of the matrix elements of $\bm\nabla\bold{\breve{H}}(\bm{q})$ in addition to those of $\bold{\breve{H}}(\bm{q})$.
Then, as shown in Appendix~\ref{app:NAC}, for $\bold{\breve{T}}(\bm{q})$ to be the inverse of the ADT (and the intermediate basis to be diabatic), it must satisfy -- as much as possible -- the following condition,
\begin{equation}
\bold{\breve{A}}(\bm{q}) = \bold{\breve{T}}^\dag(\bm{q})\bm\nabla\bold{\breve{T}}(\bm{q}) \approx \bold{\breve{F}}^\mathrm{CI}(\bm{q}) \quad,
\end{equation}
where $\bold{\breve{F}}^\mathrm{CI}(\bm{q})$ is the block-matrix restriction to the target subspace of $\bold{F}^\mathrm{CI}(\bm{q}) = \bold{C}^\dag(\bm{q})\bm\nabla\bold{C}(\bm{q})$, the complete CI-NAC matrix (see Appendix~\ref{app:NAC} for further details).
This is the `fundamental condition of diabaticity' as regards the CI-NACs within the target subspace, and is mathematically far less trivial than it seems at first sight, as discussed in detail for example in Ref. \cite{bae06}.

The situation is known to be somewhat easier to handle in the two-state case with real-valued many-body wavefunctions. 
There, we have, at regular points, 
\begin{subequations}
\begin{align}
& \bm\nabla\bold{\breve{T}}(\bm{q})
= -i \bm\nabla \varphi(\bm{q}) \bold{\breve{T}}(\bm{q}) \boldsymbol\sigma_2 \quad, \\ 
& \bold{\breve{A}}(\bm{q})
= -i \bm\nabla \varphi(\bm{q}) \boldsymbol\sigma_2 \quad,
\end{align}
\end{subequations}
such that the latter is a real-skew-symmetric matrix of vectors when it is definite, 
\begin{equation}
\bold{\breve{A}}(\bm{q})
=\begin{pmatrix}
\bm0 & -\bm\nabla \varphi(\bm{q}) \\
\bm\nabla \varphi(\bm{q}) & \bm0
\end{pmatrix}
\quad.
\end{equation}
The CI and CSF contributions to the ``residual'' and adiabatic NACs satisfy (see Appendix~\ref{app:NAC} for their general definitions), 
\begin{subequations}
\begin{align}
& \bold{\breve{f}}^\mathrm{CI}(\bm{q}) = \bold{\breve{T}}(\bm{q})
\big( \bold{\breve{F}}^\mathrm{CI}(\bm{q} ) - \bold{\breve{A}}(\bm{q}) \big)
\bold{\breve{T}}^\dag(\bm{q}) \quad, \\
& \bold{\breve{f}}^\mathrm{CSF}(\bm{q}) = \bold{\breve{T}}(\bm{q})\bold{\breve{F}}^\mathrm{CSF}(\bm{q})\bold{\breve{T}}^\dag(\bm{q}) \quad.
\end{align}
\end{subequations}
They also are real-skew-symmetric, \emph{i.e.},
\begin{subequations}
\begin{align}
& \bold{\breve{f}}^\mathrm{CI/CSF}(\bm{q})
= i \bm{f}_{01}^\mathrm{CI/CSF}(\bm{q}) \boldsymbol{\sigma}_2
\quad, \\
& \bold{\breve{F}}^\mathrm{CI/CSF}(\bm{q})
= i \bm{F}_{01}^\mathrm{CI/CSF}(\bm{q}) \boldsymbol{\sigma}_2
\quad.
\end{align}
\end{subequations}
Owing to the algebraic commutation rules of rotation matrices and real-skew-symmetric matrices, all generated by $i\boldsymbol\sigma_2$, we get the so-called Abelian (\emph{i.e.}, without matrix) vector formulation of the problem (see, \emph{e.g.}, Ref. \cite{bae06} and references therein),
\begin{subequations}
\begin{align}\label{eq:NAC_dia_adia}
& \bm{f}_{01}^\mathrm{CI}(\bm{q}) = \bm{F}_{01}^\mathrm{CI}(\bm{q}) + \bm\nabla \varphi(\bm{q}) \quad, \\
& \bm{f}_{01}^\mathrm{CSF}(\bm{q}) = \bm{F}_{01}^\mathrm{CSF}(\bm{q}) \quad.
\end{align}
\end{subequations}
Interestingly enough, both CSF-NACs are identical here.

As is well-known, the criterion of diabaticity for two states,
\begin{equation}
\bm{f}_{01}(\bm{q}) = \bm{f}_{01}^\mathrm{CI}(\bm{q}) + \bm{f}_{01}^\mathrm{CSF}(\bm{q}) \approx \bm0 \quad,
\end{equation}
where both contributions should be negligible, thus boils down to two adiabatic conditions,
\begin{subequations}
\begin{align}
& \bm{F}_{01}^\mathrm{CI}(\bm{q}) \approx -\bm\nabla \varphi(\bm{q}) \quad, \\
& \bm{F}_{01}^\mathrm{CSF}(\bm{q}) \approx \bm0 \quad.
\end{align}
\end{subequations}
The values of the CI and CSF contributions to the NAC in the adiabatic basis can be computed from analytic derivative techniques (see Ref. \cite{yal22:776}). 
The CSF contribution is a nonremovable part of the NAC but the extent of its magnitude largely depends on the choice of underlying orbitals and their own diabaticity (see Appendix~\ref{app:diaborbs}).
The gradient of the ADT angle is the removable part of the NAC (see Appendix~\ref{app:NAC}).
It can be obtained for example from Eq.~(\ref{eq:grads-nacs}).

As already mentioned, the criterion $\bm{F}_{01}^\mathrm{CI}(\bm{q}) \approx -\bm\nabla \varphi(\bm{q})$ provides the historical \emph{a posteriori} avenue for determining $\varphi(\bm{q})$ from the knowledge of $\bm{F}_{01}^\mathrm{CI}(\bm{q})$ upon line-integration from a reference point, which can even be extended in matrix form to more than two states (see Ref. \cite{bae06} and, \emph{e.g.}, Ref. \cite{ric20:154108} for a recent implementation in the context of on-the-fly quantum dynamics). 
In the present work, we expect an \emph{a priori} realization of this condition, upon bock-diagonalization, according to the least-transformation principle such as prescribed by Cederbaum \emph{et al.} in Refs. \cite{pac88:7367,ced89:2427}.

\subsection{Relation with \emph{ab initio} optimal diabatization schemes}\label{subsec:diab}

Now that we have introduced the formal tools for describing our two-state model and target subspaces as well as their various NAC contributions, the natural question is: \textit{how can we get an optimal diabatic basis from those?}
Its construction is non-trivial in general and can be  explored through a least-transformation principle, as introduced formally by Cederbaum \emph{et al.}, and later re-explored by Werner \emph{et al.} in the practical context of CI and SA-MCSCF methods.
We refer the interested reader to Appendix~\ref{app:NAC} where we introduce more details on such seminal works.

For now, we summarize some key aspects of these ideas that will be relevant to the next Subsec.~\ref{subsec:descriptors} where we introduce two new descriptors of diabaticity.
These provide further insight into the assessment of (quasi-)diabatic states within our approach, in which we consider the SA-VQE algorithm as an electronic-structure method that naturally converges toward a solution that is close enough to the optimal diabatic one but without requiring any post-processing transformation.

The block-diagonalization transformation from $\{\ket{\Phi^0_J;\bm{q}}\}_{J=0}^{N-1}$ to $\{\ket{\Phi_J;\bm{q}}\}_{J=0}^{N-1}$ is represented with a real-orthogonal matrix, $\bold{O}(\bm{q})$, in the real-state case. 
Its $(2 \times 2)$-restriction to the target subspace, denoted $\bold{\breve{O}}(\bm{q})$ is not an orthogonal matrix but is expected to be as close as possible to the $(2 \times 2)$-identity matrix for it to convey as much diabaticity as possible from the model subspace to the target subspace (see Appendix~\ref{app:lowdin}).

Assuming knowledge of the first two adiabatic CI-coefficient column-vectors within $\bold{C}(\bm{q})$ [see Eq.~(\ref{eq:cmatstruct})] and the existence of an ADT matrix $\bold{\breve{T}}^\dag(\bm{q})$ (the post-variational objective), we can specify the two block-diagonalizing CI-coefficient column-vectors in terms of $(N \times 2)$-matrices as
\begin{equation}
\begin{pmatrix}
\bold{\breve{O}}(\bm{q}) \\
\bold{X}(\bm{q})
\end{pmatrix}
=
\begin{pmatrix}
\bold{\breve{C}}(\bm{q})\bold{\breve{T}}^\dag(\bm{q}) \\
\bold{X}(\bm{q})
\end{pmatrix} \quad.
\end{equation}

The least-transformation criterion for ``optimal diabatization'' (further denoted with a five-branch star index when being reached) can be viewed as achieving the following condition as much as possible at any $\bm{q}$,
\begin{equation}
\label{eq:proxid}
\begin{pmatrix}
\bold{\breve{O}}_\star(\bm{q}) \\
\bold{X}_\star(\bm{q})
\end{pmatrix}
\approx
\begin{pmatrix}
\bm1 \\
\bm0
\end{pmatrix} \quad.
\end{equation}
In short, this means that the model states are ``best guesses'' everywhere for the block-diagonalization procedure: $\{\ket{\Phi_\rmA;\bm{q}},\ket{\Phi_\rmB;\bm{q}}\} \approx \{\ket{\Phi^0_\rmA;\bm{q}},\ket{\Phi^0_\rmB;\bm{q}}\}$.

It must be understood that this now relies on the assumption that the CSFs themselves can be made ``as diabatic as possible'' through the underlying set of molecular orbitals.
This corresponds to the concept of `diabatic orbitals', such as implemented in the \textsc{Molpro} quantum-chemistry software \cite{wer88:3139,sim99:4523} (see also Appendix~\ref{app:diaborbs} for further details). 

More precisely, the optimal ADT matrix should satisfy
\begin{equation}
\label{eq:T_Lowdin}
\bold{\breve{T}}^\dag_\star(\bm{q})=\bold{\breve{C}}^\dag(\bm{q}) \left(\bold{\breve{C}}(\bm{q})\bold{\breve{C}}^\dag(\bm{q}) \right)^{-1/2} \quad. 
\end{equation}
Quite obviously, the previous construction is a Löwdin Hermitian orthonormalization of $\bold{\breve{C}}^\dag(\bm{q})$ such that $\bold{\breve{T}}^\dag_\star(\bm{q})$ is the closest unitary transformation to it (see Appendix~\ref{app:lowdin}).
It is equivalent to assuming that the optimal quasi-diabatic states are such that
\begin{equation}
\bm1=\bold{\breve{O}}^\dag_\star(\bm{q})\left(\bold{\breve{O}}_\star(\bm{q})\bold{\breve{O}}^\dag_\star(\bm{q}) \right)^{-1/2} \quad,
\end{equation}
\emph{i.e.},
\begin{equation}
\bold{\breve{O}}_\star(\bm{q})=\bold{\breve{S}}^{1/2}(\bm{q})=\bold{\breve{O}}^{\dag}_\star(\bm{q}) \quad,
\end{equation}
where
\begin{equation}
\bold{\breve{S}}(\bm{q})=\bold{\breve{C}}(\bm{q})\bold{\breve{C}}^\dag(\bm{q})=\bold{\breve{O}}_\star(\bm{q})\bold{\breve{O}}^\dag_\star(\bm{q}) \quad,
\end{equation}
is an invariant Gram matrix within the target subspace and represents an intrinsic measure of the quality of the model as a guess with respect to the target. 
This implies that the optimal $\bold{\breve{O}}_\star(\bm{q})$-matrix is nonunitary but Hermitian and is as close as possible to the identity matrix.

\subsection{Descriptors of diabaticity within a two-state target subspace}\label{subsec:descriptors}

Physically speaking, the optimal diabatic states are the orthonormal and least-transformed projections of the orthonormal model states to the nonparallel target subspace. 
It must be understood here that such an approach is meant to provide an optimal ADT matrix $\bold{\breve{T}}^\dag_\star(\bm{q})$ from the knowledge of $\bold{\breve{C}}(\bm{q})$ and such that $\bold{\breve{O}}_\star(\bm{q})=\bold{\breve{C}}(\bm{q})\bold{\breve{T}}^\dag_\star(\bm{q})$.

In the present work, the situation is different: we know all matrices and the $\bold{\breve{C}}(\bm{q})$-matrix is supposed to be the same as above, but we want to check the extent to which our actual $\bold{\breve{O}}(\bm{q})$-matrix and $\bold{\breve{T}}(\bm{q})$-matrix satisfy such optimal requirements. Surprisingly enough, the most important relation of the present work is perhaps to be found in Eq.~(\ref{eq:proxid}). 
We are not looking to enforce this criterion but we trust the SA-VQE algorithm to be ``as lazy as possible'' (least-transforming) when optimizing the variational $\bmtheta(\bm{q})$-parameters (see Sec. \ref{sec:compdet}) for this to hold well numerically when achieving the block-diagonalization (variational objective). 

Now, let us examine how such considerations manifest themselves in the two-state case examined in the present work. First, we have (square SVD; see Appendix~\ref{app:lowdin}) 
\begin{equation}
\bold{\breve{O}}(\bm{q}) = \bold{\breve{U}}(\bm{q})  \bold{\breve{\Sigma}}(\bm{q})\bold{\breve{W}}^\dag(\bm{q}) \quad,
\end{equation}
and (reverse polar decomposition; see also  Appendix~\ref{app:lowdin})
\begin{equation}
\bold{\breve{O}}(\bm{q}) = \bold{\breve{O}}_\star(\bm{q})\bold{\breve{B}}(\bm{q}) \quad,
\end{equation}
where
\begin{subequations}
\begin{align}
&\bold{\breve{O}}_\star(\bm{q}) = \bold{\breve{U}}(\bm{q}) \bold{\breve{\Sigma}}(\bm{q}) \bold{\breve{U}}^\dag(\bm{q}) \quad, \\
&\bold{\breve{B}}(\bm{q}) = \bold{\breve{U}}(\bm{q}) \bold{\breve{W}}^\dag(\bm{q}) \quad.
\end{align}
\end{subequations}
We observed in the present work that our numerical SVD algorithm occurs to provide matrices for the left and right singular vectors that are real-symmetric and correspond to orthogonal rotoreflections,
\begin{align}
\label{eq:Umatrix}
\bold{\breve{U}}(\bm{q}) &= \bold{\breve{U}}^\dag(\bm{q}) = \bold{\breve{U}}^{-1}(\bm{q})
=\begin{pmatrix}
-\cos{u(\bm{q})} & \sin{u(\bm{q})} \\
\sin{u(\bm{q})} & \cos{u(\bm{q})}
\end{pmatrix} \\ \nonumber
&= -\boldsymbol\sigma_3 e^{-iu(\bm{q})\boldsymbol\sigma_2}
= -e^{iu(\bm{q})\boldsymbol\sigma_2} \boldsymbol\sigma_3
\quad,
\end{align}
and
\begin{align}
\label{eq:Wmatrix}
\bold{\breve{W}}(\bm{q}) &= \bold{\breve{W}}^\dag(\bm{q}) = \bold{\breve{W}}^{-1}(\bm{q})
=\begin{pmatrix}
-\cos{w(\bm{q})} & \sin{w(\bm{q})} \\
\sin{w(\bm{q})} & \cos{w(\bm{q})}
\end{pmatrix} \\ \nonumber
&= -\boldsymbol\sigma_3 e^{-iw(\bm{q})\boldsymbol\sigma_2}
= -e^{iw(\bm{q})\boldsymbol\sigma_2} \boldsymbol\sigma_3 
\quad,
\end{align}
where both $u(\bm{q})$ and $w(\bm{q})$ are angles supposed to have smooth and regular behaviors with respect to $\bm{q}$ and taking mutually similar values close to zero.

Then, the nonincreasing-ordered and positive definite singular-value matrix can be expressed as 
\begin{align}
\bold{\breve{\Sigma}}(\bm{q})
&=\begin{pmatrix}
\Sigma_\rmA(\bm{q}) & 0 \\
0 & \Sigma_\rmB(\bm{q})
\end{pmatrix} \\ \nonumber
&= \Sigma_+(\bm{q}) \bm1
+ \Sigma_-(\bm{q}) \boldsymbol\sigma_3
\quad,
\end{align}
where $1 \geq \Sigma_\rmA(\bm{q}) \geq \Sigma_\rmB(\bm{q}) \geq 0$ and
\begin{subequations}
\begin{align}
&0 \leq \Sigma_+(\bm{q}) = \frac{\Sigma_\rmA(\bm{q})+\Sigma_\rmB(\bm{q})}{2} \approx 1 \quad, \\
&0 \leq \Sigma_-(\bm{q}) = \frac{\Sigma_\rmA(\bm{q})-\Sigma_\rmB(\bm{q})}{2} \approx 0 \quad.
\end{align}
\end{subequations}

Hence, thanks to commutation rules, we have
\begin{subequations}
\begin{align}
&\bold{\breve{O}}_\star(\bm{q}) = e^{iu(\bm{q})\boldsymbol\sigma_2} \bold{\breve{\Sigma}}(\bm{q}) e^{-iu(\bm{q})\boldsymbol\sigma_2} \quad, \\
&\bold{\breve{B}}(\bm{q}) = e^{i(u(\bm{q})-w(\bm{q}))\boldsymbol\sigma_2} \quad.
\end{align}
\end{subequations}
Note that $-\boldsymbol\sigma_3$ has been absorbed by $\bold{\breve{\Sigma}}(\bm{q})$ on both sides, owing to the product properties of Pauli matrices.
Then,
\begin{equation}
\bold{\breve{O}}_\star(\bm{q})
= \Sigma_+(\bm{q}) \bm1
+ \Sigma_-(\bm{q}) \boldsymbol\sigma_3 e^{-i2u(\bm{q})\boldsymbol\sigma_2} \quad,
\end{equation}
while
\begin{align}
\bold{\breve{O}}(\bm{q})
&= \Sigma_+(\bm{q}) e^{i(u(\bm{q})-w(\bm{q}))\boldsymbol\sigma_2} \\ \nonumber
&+ \Sigma_-(\bm{q}) \boldsymbol\sigma_3 e^{-i(u(\bm{q})+w(\bm{q}))\boldsymbol\sigma_2} \quad.
\end{align}
Quite obviously, $\bold{\breve{O}}(\bm{q})$ identifies to $\bold{\breve{O}}_\star(\bm{q})$ when $w(\bm{q}) \equiv u(\bm{q})$; their difference is thus an objective measure of the deviation with respect to optimal diabaticity.

It is illuminating to further consider the following two descriptors that we introduce here, based on matrix Frobenius norms.
First, we can evaluate the intrinsic optimality of the diabatization as follows,
\begin{align}
\label{eq:dstar}
d(\bm{q})
&= \norm{\bold{\breve{C}}(\bm{q})-\bold{\breve{T}}_\star(\bm{q})} \\ \nonumber
&= \norm{\bold{\breve{O}}_\star(\bm{q})-\bm1} \\ \nonumber
&= \norm{\bold{\breve{\Sigma}}(\bm{q})-\bm1} \\ \nonumber
&= \sqrt{(\Sigma_\rmA(\bm{q})-1)^2+(\Sigma_\rmB(\bm{q})-1)^2} \quad.
\end{align}
It tells us how much the model and target subspaces match (CSF contribution to the NACs).
Second, we can evaluate the residual deviation of our actual block-diagonalization with respect to optimality as follows,
\begin{align}
\label{eq:rstar}
r(\bm{q})
&= \norm{\bold{\breve{B}}(\bm{q})-\bm1} \\ \nonumber
&= \norm{\bold{\breve{U}}(\bm{q}) - \bold{\breve{W}}(\bm{q})} \\ \nonumber
&= 2 \sqrt{1-\cos{(u(\bm{q})-w(\bm{q})})} \quad.
\end{align}
Among other things, it tells us how close $\bold{\breve{O}}(\bm{q})$ is to being Hermitian and thus optimal as regards diabaticity (CI contribution to the NACs). 

The two prominent descriptors identified above are directly related to the geometric properties of the square SVD. 
A variety of complementary descriptors can be defined (\emph{e.g.}, $\norm{\bold{\breve{O}}(\bm{q})-\bold{\breve{O}}_\star(\bm{q})}$, $\norm{\bold{\breve{O}}(\bm{q})-\bm1}$, or $\norm{\bold{\breve{O}}(\bm{q})-\bold{\breve{O}}^\dag(\bm{q})}$), but they will only ever mix together and convey the same type of information.
Also, let us stress here that the primary definitions of our two descriptors are general, based on submatrix Frobenius norms, and do not rely on the limiting case of a two-state model.

\section{Method: State-average orbital-optimized variational quantum eigensolver}\label{sec:compdet}

As aforementioned, the SA-OO-VQE algorithm is the quantum analog of SA-MCSCF and can be viewed, as such, as the natural extension of VQE~\cite{peruzzo2014variational} to excited states~\cite{nakanishi2019subspace,yal21:024004,yal22:776}. 
Based on the ensemble extension of the Rayleigh--Ritz variational principle~\cite{theophilou1979energy,gross1988rayleigh,ding2024ground},
SA-OO-VQE is a hybrid quantum/classical algorithm designed to compute the ensemble energy, a
weighted sum of the ground- and excited-state energies
of a given Hamiltonian $\hat{H}^{\rm el}[\bmq]$, usually written as a linear combination of Pauli strings within a QC context.

For convenience, only two states will be considered herein, although a generalization to larger ensembles is straightforward.
In this case, the goal (variational objective) of SA-OO-VQE is to determine
the subspace spanned by $\lbrace \ket{\Psi_0;\bmq},\ket{\Psi_1;\bmq} \rbrace$.
For the sake of convenience, we shall get rid of the dependence on $\bmq$ in the present section, as we do not have to consider two different geometries.
The algorithm can be divided into the following steps, where the orbital optimization (OO) will be considered later on for simplicity:
\begin{enumerate}
    \item Initialization. 
    The first step consists in choosing simple-to-prepare and chemically intuitive orthonormal initial states
    $\lbrace \ket{\Phi_{\rm A}^0}, \ket{\Phi_{\rm B}^0}\rbrace$ (the model, or guess).
    In particular, the closed-shell Hartree--Fock-type SD and the singly-excited $n\pi^*$-type singlet CSF of formaldimine are considered in the present work (see below), according to previous ones~\cite{yal21:024004,yal22:776} (herein, in the basis of active diabatic orbitals). 
    Hence, the number of different quantum circuits corresponds to the number of states of the ensemble (here: two).
    \item State preparation. 
    A parametrized unitary operator (determining the ansatz of the many-body method; here: the trotterized -- also called disentangled -- GUCCSD) $\hat{U}_{\rm SA-VQE}(\bmtheta)$ is applied to each initial state, thus leading to
    $\ket{\Phi_{\rm A}(\bmtheta)} = \hat{U}_{\rm SA-VQE}(\bmtheta) \ket{\Phi_{\rm A}^0}$
    and $\ket{\Phi_{\rm B}(\bmtheta)} = \hat{U}_{\rm SA-VQE}(\bmtheta) \ket{\Phi_{\rm B}^0}$ (the target, or variational objective at stationarity, \emph{i.e.},  convergence).
    Since this is a unitary transformation, the orthogonality between $\ket{\Phi_{\rm A}(\bmtheta)}$ and $\ket{\Phi_{\rm B}(\bmtheta)}$ is conserved.
    \item Quantum measurements. Statistical sampling of the parametrized quantum circuits is performed by a series of shots by the quantum computer (or deterministically from its state-vector classical emulator, as in here) so as to estimate the probabilistic expectation values of the Hamiltonian with respect to the prepared states, denoted $H_{\rm AA}(\bmtheta) = \bra{\Phi_{\rm A}(\bmtheta)} \hat{H}^{\rm el}\ket{\Phi_{\rm A}(\bmtheta)}$
    and $H_{\rm BB}(\bmtheta) = \bra{\Phi_{\rm B}(\bmtheta)} \hat{H}^{\rm el}\ket{\Phi_{\rm B}(\bmtheta)}$.
    \item Classical optimization. The previous steps are meant to have been performed on a quantum computer for a fixed set of parameters $\bmtheta$, sometimes initialized to zero, or to random numbers. 
    These parameters are then to be further optimized thanks to an interfaced classical computer, according to the ensemble variational principle,
\begin{eqnarray}\label{eq:var_ES}
    E^{\rm SA}(\bmtheta^*)&=& \min_\bmtheta\lbrace  H_{\rm AA}(\bmtheta)+H_{\rm BB}(\bmtheta) \rbrace\nonumber \\ &\geq& E^{\rm SA}
    =\mathcal{E}_0+\mathcal{E}_1 \quad,
\end{eqnarray}
    where $E^{\rm SA}(\bmtheta^*)$ refers to the state-average energy that is minimized by the optimal set of parameters $\bmtheta^*$.
    In this, $\mathcal{E}_0$ and $\mathcal{E}_1$ are
    the energies defined in Eq.~(\ref{eq:cieigen}), \textit{i.e.}, the lowest two eigenvalues of the Hamiltonian $\hat{H}^{\rm el}$.
    Depending on the calculation of energy gradients or not, gradient-free or gradient-based classical optimizers can be used. 
    The choice of an efficient classical optimizer is a very active field of research for variational quantum algorithms~\cite{bonet2023performance}.
    The new set of parameters determined by the classical optimizer is then plugged into the parametrized quantum circuit in step 2, and so on until SA-VQE convergence is reached with respect to $\bmtheta$.
\end{enumerate}

Quite importantly, since the SA energy is invariant under any rotation (or any type of unitary transformation) among the target states, its variational minimization only ensures the actual block-diagonalization of $\hat{H}^{\rm el}$ (true variational objective).
This brings a crucial difference with respect to the `subspace search VQE with different weights', such as exposed in Ref.~\cite{nakanishi2019subspace},
which ensures that the states
$\ket{\Phi_{\rm A}(\bmtheta^*)}$ and $\ket{\Phi_{\rm B}(\bmtheta^*)}$ should be
as close as possible to the eigenstates $\ket{\Psi_0}$
and $\ket{\Psi_1}$, and
$H_{\rm AA}(\bmtheta^*)$ and $H_{\rm BB}(\bmtheta^*)$ to $\mathcal{E}_0$ and $\mathcal{E}_1$, respectively.

In any case, the eigenstates, $\ket{\Psi_0}$
and $\ket{\Psi_1}$,
can always be obtained \textit{a posteriori} within the equal-weighted SA approach upon performing a final rotation of the converged target states, $\lbrace \ket{\Phi_{\rm A}(\bmtheta^*)},\ket{\Phi_{\rm B}(\bmtheta^*)}\rbrace$.
Such a rotation is defined through
the post-variational diagonalization angle $\varphi$, as shown in Eq.~(\ref{eq:rotation}).
This specific rotation can be executed on a dedicated quantum circuit according to Ref.~\cite{yal22:776} so as to prepare $\ket{\Psi_0}$
and $\ket{\Psi_1}$ on the quantum computer and have them available for further probing.

Finally, when only a limited part of the Hilbert space is considered,
\emph{e.g.}, in the case of an active-space selection, the Hamiltonian spectrum is not invariant with respect to orbital rotations anymore.
Let us denote such a Hamiltonian the frozen-core Hamiltonian $\hat{H}^{\rm FC}$ (see Ref.~\cite{yal21:024004} for more details on its construction).
In this case, although the prepared
states are variational with respect to the circuit parameters, they are not with respect to
the orbitals, which can be optimized on a classical computer upon applying the unitary OO operator,
\begin{equation}
    \hat{U}_{\rm OO}(\bmkappa) = e^{-\hat{\kappa}} \quad,
\end{equation}
in order to obtain the following unitary similarity transform of the frozen-core Hamiltonian,
\begin{equation}
   \hat{H}^{\rm FC}(\bmkappa) = \hat{U}_{\rm OO}^\dagger (\bmkappa) \hat{H}^{\rm FC} \hat{U}_{\rm OO}(\bmkappa) \quad,
\end{equation}
with
\begin{equation}
    \hat{\kappa} = \sum_{p>q} \sum_\sigma \kappa_{pq}(\hat{a}_{p\sigma}^\dagger\hat{a}_{q\sigma} - \hat{a}^\dagger_{q\sigma} \hat{a}_{p\sigma}) \quad,
\end{equation}
where $p$ and $q$ denote spatial orbitals and $\sigma = \lbrace \uparrow, \downarrow \rbrace$ the spin.
Hence, the optimal OO parameters, $\bmkappa^*$,
are found variationally upon minimizing the SA energy,
\begin{eqnarray}\label{eq:var_ES}
    E^{\rm SA}(\bmtheta^*,\bmkappa^*)&=& \min_\bmkappa\min_\bmtheta\lbrace  H_{\rm AA}(\bmtheta,\bmkappa)+H_{\rm BB}(\bmtheta,\bmkappa) \rbrace \quad,\nonumber\\
\end{eqnarray}
where $H_{\rm AA}(\bmtheta,\bmkappa) = \bra{\Phi_{\rm A}(\bmtheta)} \hat{H}^{\rm FC}(\bmkappa) \ket{\Phi_{\rm A}(\bmtheta)}$
and
$H_{\rm BB}(\bmtheta,\bmkappa) = \bra{\Phi_{\rm B}(\bmtheta)} \hat{H}^{\rm FC}(\bmkappa)\ket{\Phi_{\rm B}(\bmtheta)}$.
The expectation value of the spin operator $\hat{S}^2$ was added as a penalty term to Eq.~(\ref{eq:var_ES}) in order to favor singlet states ($\langle \hat{S}^2 \rangle = 0$), as shown in the constrained VQE algorithm~\cite{ryabinkin2018constrained}. Hence, SA-OO-VQE consists of
two variational minimization processes, one for the circuit parameters (SA-VQE; $\bmtheta$) and one for the orbitals (OO; $\bmkappa$), which are to be performed sequentially until global SA-MCSCF-type convergence is reached.

Let us remark that the generalization of the Rayleigh--Ritz variational principle to an ensemble of states has subsequently been considered in ADAPT-VQE~\cite{grimsley2025challenging} and the contracted quantum eigensolver~\cite{benavides2024quantum} to extract excited states.
Besides, as an alternative to what is presented in this section, one could target all the states in the ensemble using a single quantum circuit using the purification technique described in Refs.~\cite{xu2023concurrent} and \cite{hong2024refining}, which requires ancilla qubits.

\section{Model system, results, and discussion}
\label{sec:resdisc}

\begin{figure}
    \centering
    \resizebox{\columnwidth}{!}{
    \includegraphics{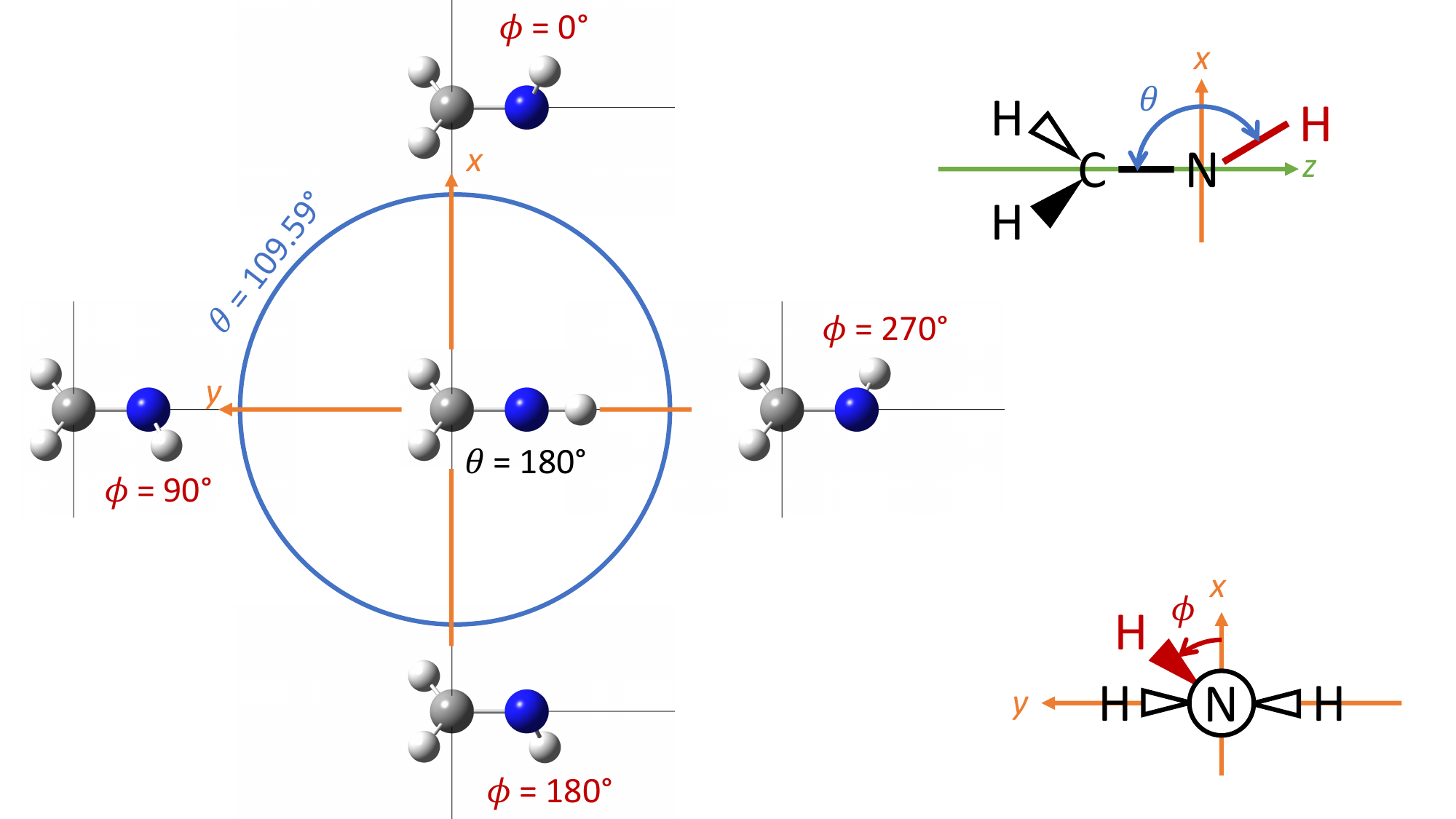}
    }
    \caption{Schematic representation of the two angular spherical coordinates and Cartesian frame defining the geometry exploration of the formaldimine molecule in this work.}
    \label{fig:geogeo}
\end{figure}

\begin{figure*}
    \centering
    \resizebox{\textwidth}{!}{
    \includegraphics{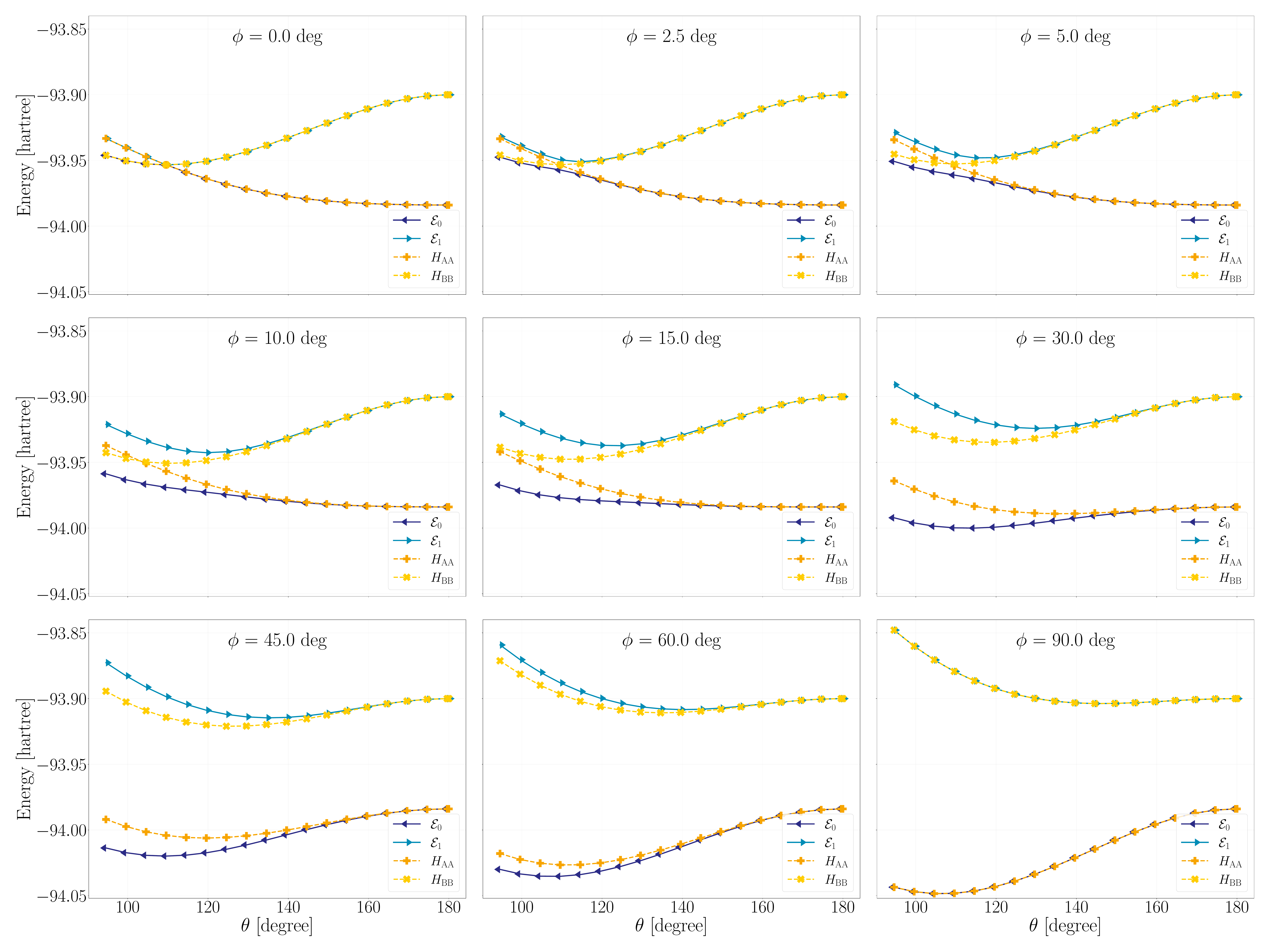}
    }
    \caption{Ground- and excited-state energy maps in the adiabatic (dark and light blue dotted lines) and diabatic (orange and yellow dotted lines) representations.}
    \label{fig:energies}
\end{figure*}

\subsection{Model system and molecular symmetry considerations}\label{sec:modsys}

\begin{figure}
    \centering
    \resizebox{\columnwidth}{!}{
    \includegraphics{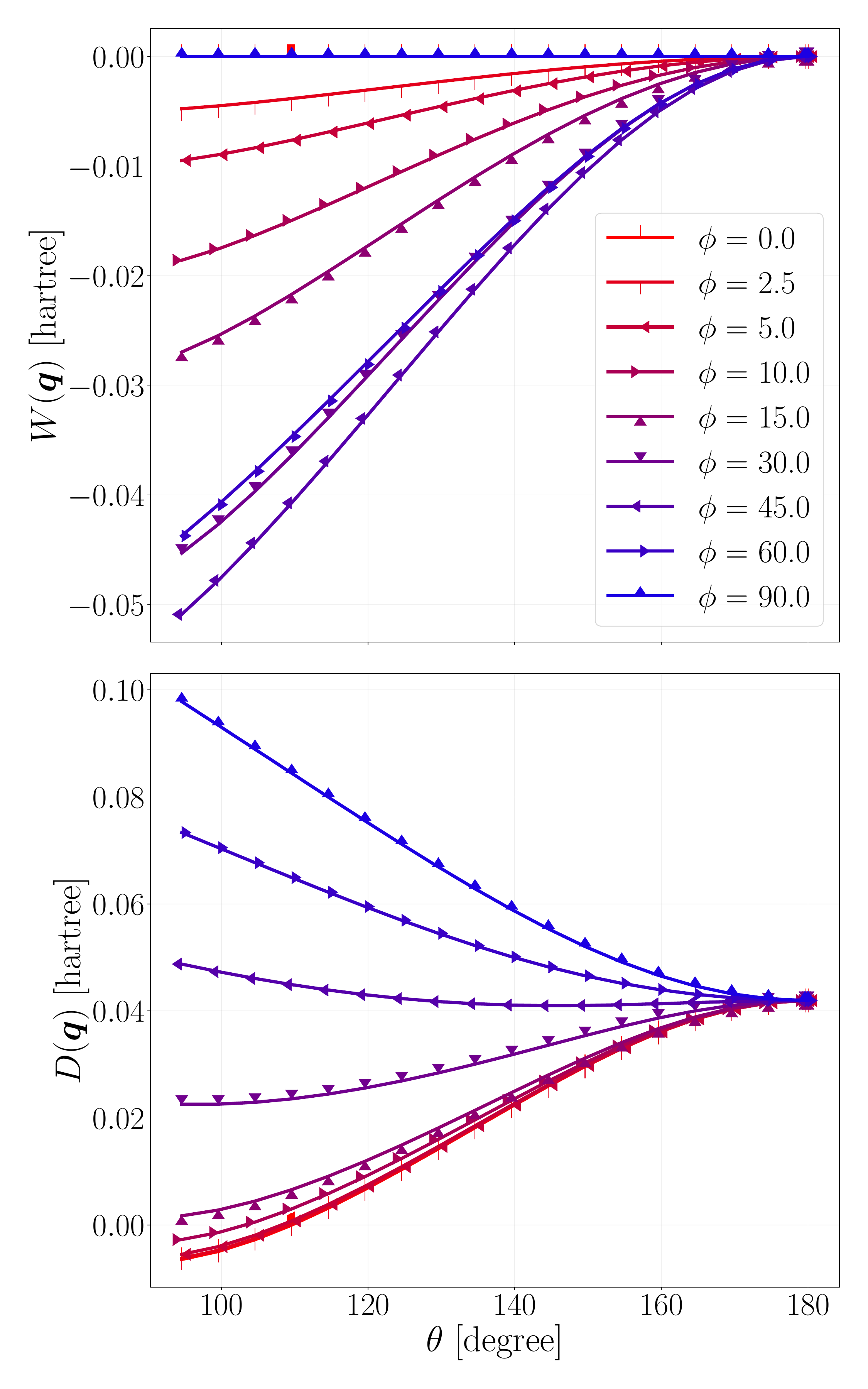}
    }
    \caption{Top panel: off-diagonal coupling element in the diabatic representation, $W(\bmq) = H_{\rm AB}(\bmq) = H_{\rm AB}(\bmq)$. Note that $W(\bmq)=0$ for both $\phi=0$ and $\phi=90$ degree. Bottom panel: diagonal diabatic entries, $D(\bmq)=(H_{\rm BB}(\bmq) - H_{\rm AA}(\bmq))/2$.
    }
    \label{fig:WD}
\end{figure}

Our ``toy-model system'' here is formaldimine, $\mathrm{H_2C=\overline{N}H}$, which is a well-known, prototypical example of a small organic molecule with a conical intersection between its ground and first-excited 
singlet electronic states, $\rm S_0$ and $\rm S_1$. 
It has been investigated in previous works by some of the present authors within a similar context \cite{yal21:024004,yal22:776}.
The description of conical intersections using quantum algorithms has also been investigated
by other groups, see for instance Refs.~\cite{koridon2024hybrid,wang2024characterizing,wang2024quantum}.

Around its ground-state equilibrium geometry, the $\rm S_0$ state is essentially a closed-shell SD of Hartree--Fock-type, while the $\rm S_1$ state is dominated by a singlet open-shell CSF of $n\pi^*$-type. 
Accounting for relevant excitations and minimal static correlation implies defining the active-orbital subset $(\pi,n,\pi^*)$ within a CAS(4,3) description (complete active space with four electrons in three spin-restricted spatial orbitals). 

The $\rm S_0$ equilibrium geometry (Franck-Condon point) is $\mathrm{C_s}$ with an in-plane conformation; there are two dynamical versions of it (terminal $\mathrm{H}$-atom, bound to $\mathrm{N}$, on 
the right or left side). 
They are connected via an in-plane $\mathrm{CNH}$ bending motion passing through a higher-symmetrical $\mathrm{C_{2v}}$ transition state (first-order saddle point) in $\rm S_0$.
There, and using Mulliken's standard conventions \cite{mul55:1997}, the $z$-axis is the principal rotation one ($\rm C_2$), and the $x$-axis is perpendicular to the molecular plane.
The $\pi$ and $\pi^*$ orbitals are $b_1$ ($x$-type) and the $n$ orbital is $b_2$ ($y$-type). 
Hence, the closed-shell ground state is $^1\rmA_1$ and the $n\pi^*$-type $S_1$ state is $^1\rmA_2$.

For exploring relevant geometries around a conical intersection, we shall consider variations of the two spherical-like angles, 
denoted $(\theta,\phi)$, of the terminal $\mathrm{H}$-atom (bound to $\mathrm{N}$). 
The $\theta$-angle corresponds to $\widehat{\rm{HNC}}$ (antipolar with respect to the $z$-axis that goes from C to N, \emph{i.e}, $\pi-\theta$ is the true spherical polar angle); the dihedral $\phi$-angle is the spherical azimuthal angle about the $z$-axis and originated from the $x$-axis; see Fig.~\ref{fig:geogeo}.

The values of the other internal coordinates were chosen such that our two-dimensional 
map contains a $\mathrm{C_s}$  point of conical intersection at $\theta=109.5902^\circ$ and $\phi=0^\circ$ (and its mirror image at $180^\circ$); 
this geometry was partially optimised as a minimum-energy conical intersection under the constraint that 
the $\mathrm{H_2CN}$-fragment remains planar, with the terminal $\mathrm{H}$-atom moving around within a hemisphere 
centred at $\mathrm{N}$ (the true minimum-energy conical intersection is also $\mathrm{C_s}$ but $sp^3$-pyramidalized around C; see Ref. \cite{yal22:776}). 
Such a choice was made in order for the spherical ``pole'' ($\theta = 180^\circ$ and undetermined $\phi$) to be a planar $\mathrm{C_{2v}}$ 
geometry playing the role of a high-symmetry origin, quite similar to the actual $\rm S_0$ transition state.
Relevant geometries are illustrated in Fig.~\ref{fig:geogeo} and defined in Appendix~\ref{app:coord}.

At the most-symmetrical $\mathrm{C_{2v}}$ geometry, the two electronic states of interest are $\rm S_0-{^1\rmA_1}$ and $\rm S_1-{^1\rmA_2}$. Infinitesimal variations of $\theta$ from $180^\circ$ (the ``pole'' origin) 
at specific values of $\phi$ (multiples of $90^\circ$) can be related to $\rmB_1(x)$ and $\rmB_2(y)$ rectilinear vibrations as follows,
\begin{align*}
&\rmB_1: \delta x > 0 \sim (\delta\theta,\phi=0^\circ) \quad &[\mathrm{C_{2v}} \leadsto \mathrm{C_{s}}(xz)] \quad, \\
&\rmB_2: \delta y > 0 \sim (\delta\theta,\phi=90^\circ) \quad &[\mathrm{C_{2v}} \leadsto \mathrm{C_{s}}(yz)] \quad, \\
&\rmB_1: \delta x < 0 \sim (\delta\theta,\phi=180^\circ) \quad &[\mathrm{C_{2v}} \leadsto \mathrm{C_{s}}(xz)] \quad, \\
&\rmB_2: \delta y < 0 \sim (\delta\theta,\phi=270^\circ) \quad &[\mathrm{C_{2v}} \leadsto \mathrm{C_{s}}(yz)] \quad.
\end{align*}
This can be extended to finite angular variations of $\theta$ from $180^\circ$ such 
that $\rmB_1$ and $\rmB_2$ rectilinear vibrations will implicitly couple with further ``transparent'' $\rmA_1$ rectilinear vibrations for preserving the length of the $\mathrm{NH}$-bond.

However, there is no $\rmA_2$-type vibrational mode at $\mathrm{C_{2v}}$ geometries in the formaldimine molecule (this would require an additional torsional-like degree of freedom around the $\mathrm{CN}$-bond in a larger system). 
We thus have to consider $\rmA_2(xy) = \rmB_1(x) \otimes \rmB_2(y)$ bilinear variations. Finite values of $xy$ are achieved within the four $(x,y)$-quadrants of the two-dimensional map when $\theta$ varies and $\phi$ is not a multiple of $90^\circ$.

Hence, if we decide that the first diabatic state, labeled $\rmA$, behaves as $\rmA_1$ (coincident with $\rm S_0$ at the $\mathrm{C_{2v}}$ origin) and the second, labeled $\rmB$, behaves as $\rmA_2$ 
(coincident with $\rm S_1$ at the $\mathrm{C_{2v}}$ origin), we know that the corresponding matrix elements representing the electronic Hamiltonian in this basis are functions 
of $(\theta,\phi)$ such that $H_{\rmA\rmA}(\theta,\phi)$ and $H_{\rmB\rmB}(\theta,\phi)$ are $\rmA_1$ (behaving in terms of combinations of $x^2$, $y^2$, and any function of $z$), while $H_{\rmA\rmB}(\theta,\phi)=H_{\rmB\rmA}(\theta,\phi)$ 
(assuming real-valued wavefunctions) is $\rmA_2= \rmA_1 \otimes \rmA_2$ and behaves as $xy$ within the two-dimensional map, with a fourfold change of sign, 
and two nodal lines, $H_{\rmA\rmB}(\forall \theta,\phi=0^\circ \mod{90^\circ}) \equiv 0$. 

\begin{figure}
    \centering
    \resizebox{\columnwidth}{!}{
    \includegraphics{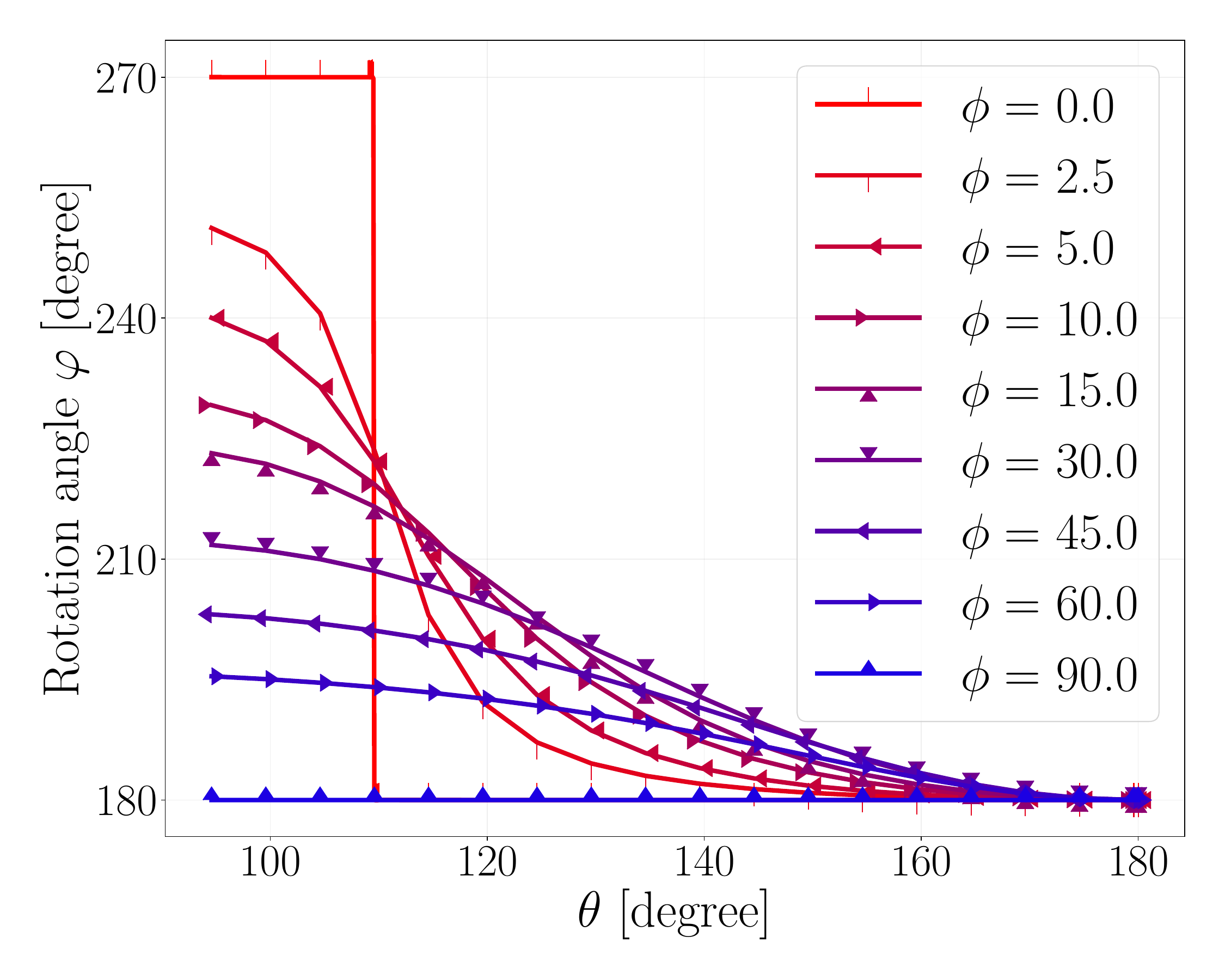}
    }
    \caption{Post-variational rotation angle of the ADT transformation from the block-diagonalization to the final partial diagonalization of the Hamiltonian $(2\times2)$-submatrix.
    }
\label{fig:varphi_singlegraph}
\end{figure}

\subsection{Results and discussion}

\begin{figure*}
    \centering
    \resizebox{\textwidth}{!}{
    \includegraphics{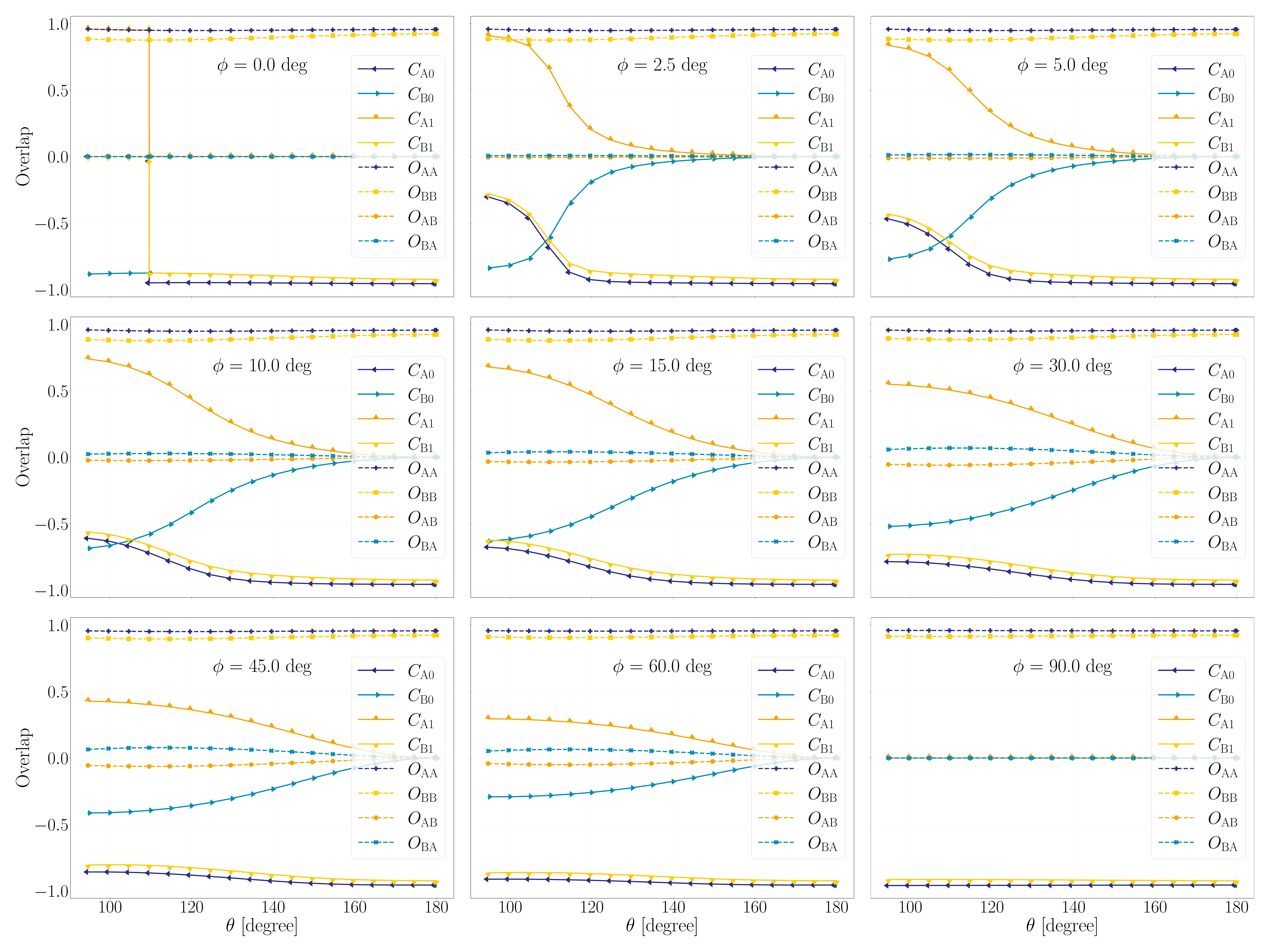}
    }
    \caption{CI-coefficient submatrix elements in the adiabatic and diabatic representations.
    }\label{fig:overlap_sign}
\end{figure*}

In what follows, we shall term ``diabatic'', for simplicity, the two aforementioned ``intermediate'' target states labeled $\rmA$ and $\rmB$, resulting from the variational CI block-diagonalization, and associated to the $\bold{\breve{O}}(\bm{q})$-coefficient and $\bold{\breve{H}}(\bm{q})$-Hamiltonian submatrices, assuming that $\bold{\breve{O}}(\bm{q})$ is close enough to the `optimally diabatic' $\bold{\breve{O}}_\star(\bm{q})$. 
The two adiabatic eigenstates $\rm S_0$ and $\rm S_1$ correspond to the $\bold{\breve{C}}(\bm{q})$-coefficient and diagonal $\bold{\breve{E}}(\bm{q})$-Hamiltonian submatrices, further obtained via the post-variational procedure (see Sec. \ref{sec:theory} for the definitions of such notations).

The diabatic and adiabatic energy maps (two-dimensional cuts through the potential energy surfaces) are represented as energy profiles along $\theta$ for various values of $\phi$ in Fig. \ref{fig:energies}.
A single hemispherical quadrant will be analyzed but it can be replicated thanks to symmetry considerations, as aforementioned.
We recall here that:
(i) there is a $\mathrm{C_s}$  point of conical intersection at $\theta=109.5902^\circ$ and $\phi=0^\circ$;
(ii) the $\mathrm{C_{2v}}$ ``pole'' origin is common to all profiles ($\theta=180^\circ$ for any $\phi$);
(iii) the geometries for which $\phi=0^\circ$ or $\phi=90^\circ$ for any $\theta$ are $\mathrm{C_s}$ (pseudopyramidal or planar, respectively) and such that $H_{\rm AB}(\bm{q}) \equiv 0$ (the adiabatic and diabatic representations thus coincide up to labelling with respect to energy ordering).

From visual inspection, the diagonal diabatic entries, $H_{\rm AA}(\bm{q})$ and $H_{\rm BB}(\bm{q})$ are smooth and regular functions of $(\theta,\phi)$ and behave as expected.
As shown in Sec. \ref{sec:theory}, the average energy, $S(\bm{q})=\frac{H_{\rmA\rmA}(\bm{q})+H_{\rmB\rmB}(\bm{q})}{2}=\frac{\mathcal{E}_{0}(\bm{q})+\mathcal{E}_{1}(\bm{q})}{2}=\mathcal{S}(\bm{q})$, is an invariant.
It clearly behaves in a smooth and regular manner and does not require any further attention here (let us remind that it is the variational objective to be minimized).
A detailed analysis of $D(\bm{q})=\frac{H_{\rmB\rmB}(\bm{q})-H_{\rmA\rmA}(\bm{q})}{2}$ and $W(\bm{q})=H_{\rmA\rmB}(\bm{q})=H_{\rmB\rmA}(\bm{q})$ is more crucial. 
Their maps are shown in Fig. \ref{fig:WD}.
Note that the only point where they both vanish together (the conical intersection) is, indeed, at $\theta=109.5902^\circ$ and $\phi=0^\circ$.

From a heuristic perspective, we can already make the assumption that the variational block-diagonalization procedure involved in SA-OO-VQE with diabatic orbitals provides ``for free'' a representation that is not far from being optimally diabatic.
This will be assessed quantitatively on more solid ground in what follows.

The maps of the ADT $\varphi(\bm{q})$-angle (implied in the final diagonalization here) are plotted in Fig. \ref{fig:varphi_singlegraph}.
This quantity can be related to the values and signs of $D(\bm{q})$ and $W(\bm{q})$ [see Eq.~(\ref{eq:polar})].
Within our selected hemispherical quadrant ($0^\circ \leq \phi \leq 90^\circ$), we observe numerically that $W(\bm{q}) \leq 0$ everywhere.
Hence, $\sin{2\varphi(\bm{q})} \geq 0$.
If we use a principal-argument angle convention (modulo $2\pi$) such that $\pi < 2\varphi(\bm{q}) \leq 3\pi$, we can consider that we are limited here to the half-domain $2\pi \leq 2\varphi(\bm{q}) \leq 3\pi$ for ensuring a nonnegative sine,
hence $\pi \leq \varphi(\bm{q}) \leq 3\pi/2$ in Fig. \ref{fig:varphi_singlegraph}. 

Further, we can discriminate two ``subsituations''.
First, we observe numerically that $D(\bm{q}) \geq 0$ almost everywhere.
Within this subdomain (including the ``
pole'' origin, $\bm{q}=\bm{q}_0$, where we can set $\varphi(\bm{q}_0)=\pi$), we have $\cos{2\varphi(\bm{q})} \geq 0$, such that $2\pi \leq 2\varphi(\bm{q}) \leq 5\pi/2$,
hence $\pi \leq \varphi(\bm{q}) \leq 5\pi/4$.
Second, there are some geometries where $D(\bm{q}) \leq 0$.
There, we have $\cos{2\varphi(\bm{q})} \leq 0$, such that $5\pi/2 \leq 2\varphi(\bm{q}) \leq 3\pi$,
hence $5\pi/4 \leq \varphi(\bm{q}) \leq 3\pi/2$.

In particular, the case for which $\phi=0^\circ$ deserves special attention because $W(\bm{q})$ is zero but $D(\bm{q})$ changes sign abruptly before and after the conical intersection, according to the value of $\theta$.
For consistency, we can consider that $D(\bm{q})>0$ ($\theta>109.5902^\circ$) corresponds to $\varphi(\bm{q})=\pi$ (as for the ``pole'' origin), while $D(\bm{q})<0$ ($\theta<109.5902^\circ$) corresponds to $\varphi(\bm{q})=3\pi/2$.
At the conical intersection, we additionally have $D(\bm{q})=0$, and the problem becomes ill-defined: the value of $\varphi(\bm{q})$ is then undetermined and lies somewhere between $\pi$ and $3\pi/2$.

Now, we can turn to the analysis of the CI-coefficient submatrices, $\bold{\breve{C}}(\bm{q})$ and $\bold{\breve{O}}(\bm{q})$.
Their maps are plotted in Fig. \ref{fig:overlap_sign}.
From visual inspection, we can observe everywhere that $\bold{\breve{O}}(\bm{q})$ is close to $\bm1$ and that $\bold{\breve{C}}(\bm{q})$ is close to $\bold{\breve{T}}(\bm{q})$ (rotation through $\varphi(\bm{q})$; see Fig. \ref{fig:varphi_singlegraph}).
In particular, at the ``pole" origin, $\bold{\breve{C}}(\bm{q})$ is close to the negative of the identity matrix, consistent with $\varphi(\bm{q})=\pi$ at this point.
It must be reminded that the absolute signs of the CI eigenvectors are formally arbitrary. 
We had to fix some of them \emph{a posteriori} in Fig. \ref{fig:overlap_sign}, consistently with the convention used for restricting the values of $\varphi(\bm{q})$ within its aforementioned principal domain considered in Fig. \ref{fig:varphi_singlegraph}.
This is a practical illustration of the well-known and critical problem of phase or sign inconsistency when using the adiabatic representation without any path-dependent constraint \cite{bae06,shu22:992,ric15:12457}. 

In contrast, we observe that $\bold{\breve{O}}(\bm{q})$ remains close to the identity matrix with consistent signs everywhere, and this being the case without any post-processing.
This tells us that the block-diagonalization realized by the SA-VQE algorithm, together with the OO and orbital diabatization procedures, truly behaves in a least-transformed manner locally, at all geometries, without further enforcing any global constraint. 
Once again, this is, in fact, a crucial result, since random phase or sign changes have been known as the main impediment for \emph{ab initio} diabatization procedures.

As discussed in Sec. \ref{sec:theory} (see also Appendix~\ref{app:lowdin}), we can evaluate quantitatively the extent of diabaticity of the diabatic representation with two descriptors built from the three matrices involved in the SVD of $\bold{\breve{O}}(\bm{q})$.
First, the intrinsic diabaticity, $d(\bm{q})$ [see Eq.~(\ref{eq:dstar})], is characteristic of the optimal diabatic representation. 
It tells us about the intrinsic deviation from maximal diabaticity due to the incompressible distance between the model and target subspaces at any geometry.
Its lower bound is zero (perfect coincidence and maximal diabaticity) and its upper bound is $\sqrt{2}$ (full mismatch).
Second, the residual diabaticity, $r(\bm{q})$ [see Eq.~(\ref{eq:rstar})], corresponds to the actual diabatic representation and is a measure of how near it is to the optimal one.
Its lower bound is zero (optimality) and its upper bound is $2\sqrt{2}$ (worst-case scenario).

\begin{figure}
    \centering
\resizebox{\columnwidth}{!}{
\includegraphics{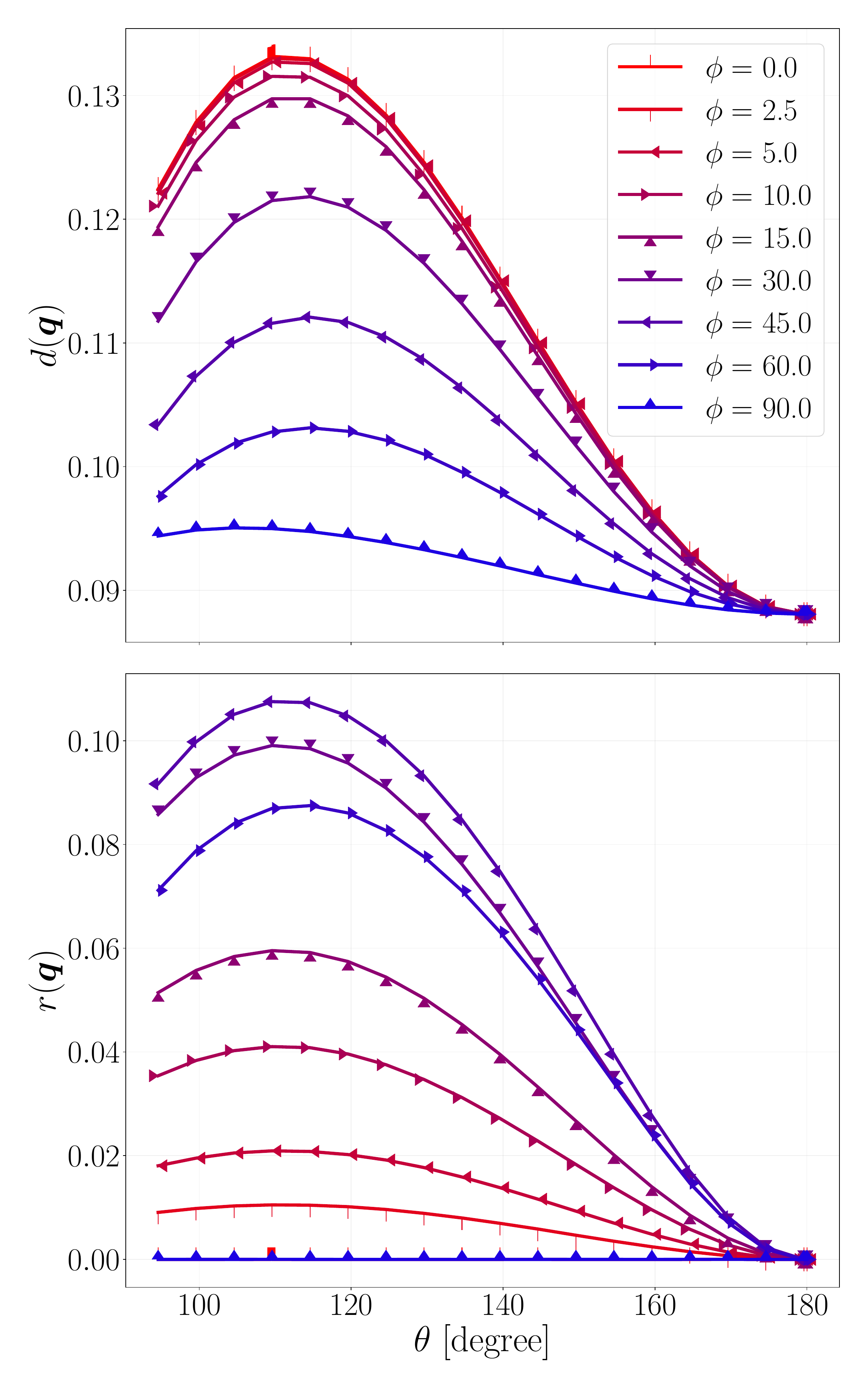}}
    \caption{Descriptors measuring the extent of diabaticity. Top panel: descriptor $d(\bmq)$, see Eq.~(\ref{eq:dstar}), which describes the distance between the model subspace and the target subspace (optimal diabaticity). Bottom panel: descriptor $r(\bmq)$, see Eq.~(\ref{eq:rstar}), which describes how far our numerically-obtained quasi-diabatic representation deviates from the optimal one. Note that $r(\bmq)=0$ for $\phi=0$ and $\phi=90$ degree for symmetry reasons.}
\label{fig:descriptor_d_and_r}
\end{figure}

\begin{figure}
    \centering
    \resizebox{0.95\columnwidth}{!}{
    \includegraphics{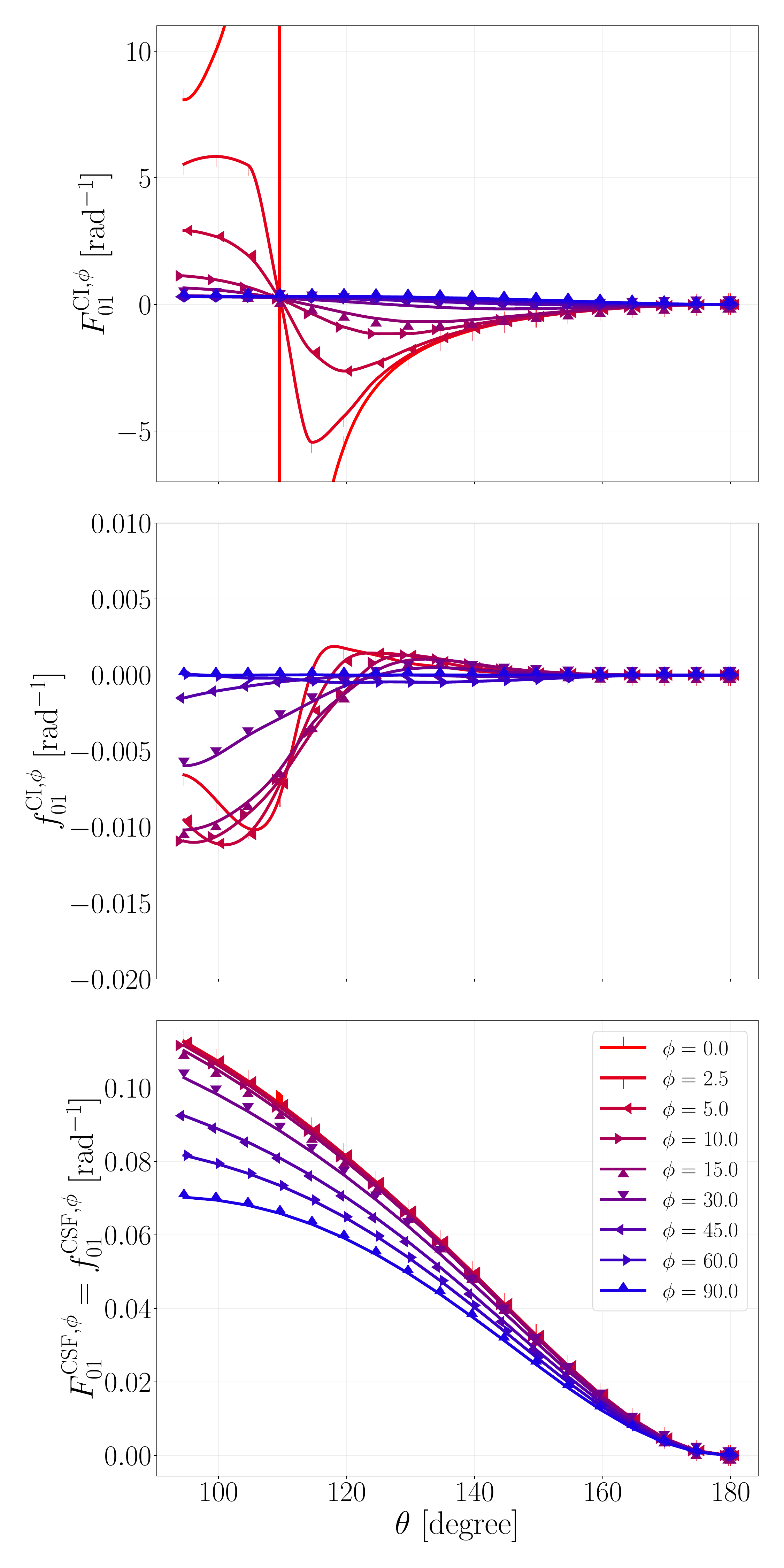} 
    }
    \caption{Component of the non-adiabatic couplings with respect to $\phi$. Top panel: CI contributions in the adiabatic representation. Middle panel: CI contributions in our quasi-diabatic representation (not shown for $\phi=0$ because they would involve ill-defined compensations between spurious numerical infinities). Spline interpolations have been used to smooth the curves. Bottom panel: CSF contributions.
    }  \label{fig:nac_phi_singleplot}
\end{figure}

\begin{figure}
    \centering
    \resizebox{0.95\columnwidth}{!}{
    \includegraphics{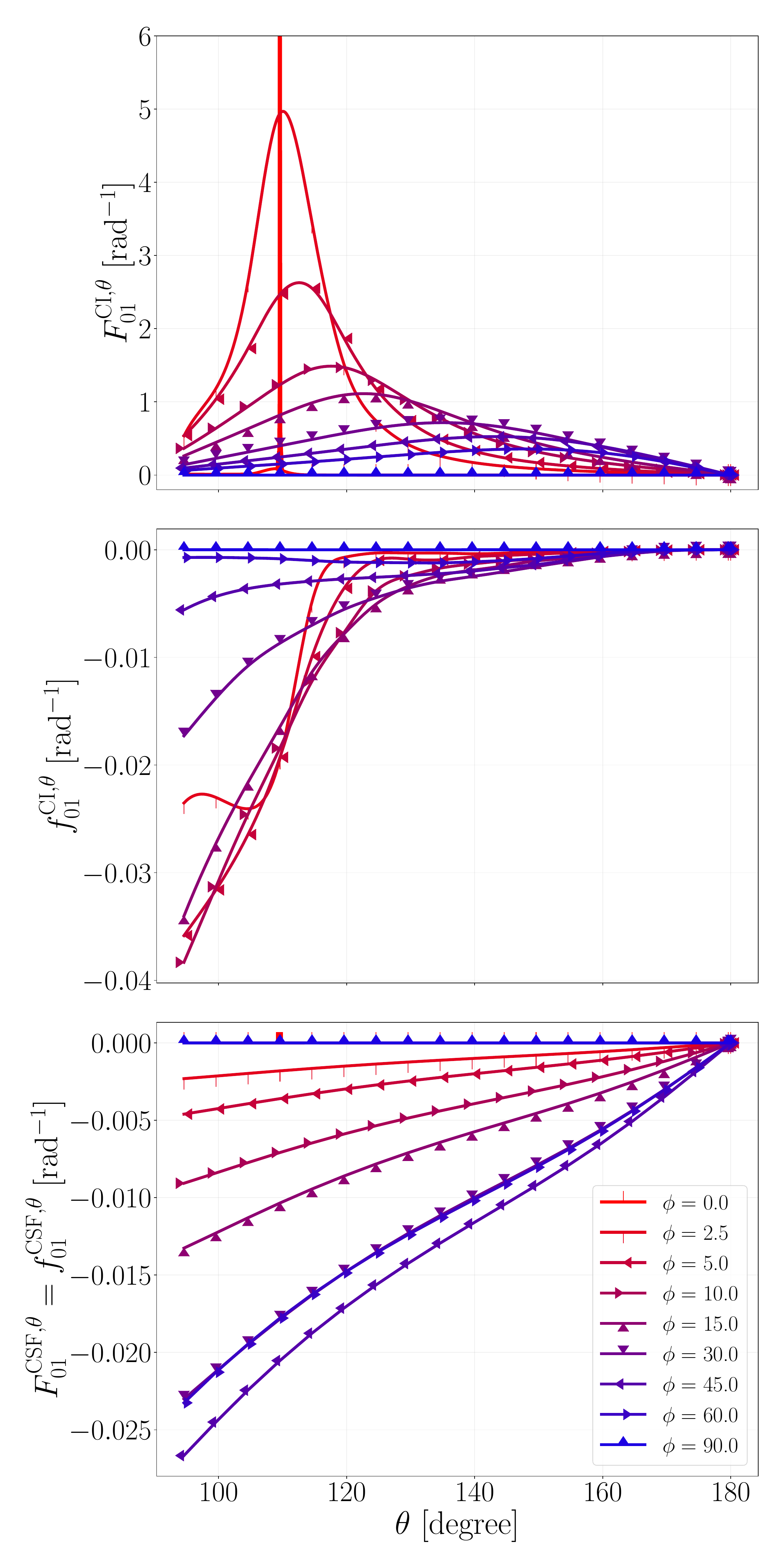} 
    }
    \caption{Component of the non-adiabatic couplings with respect to $\theta$. Top panel: CI contributions in the adiabatic representation. Middle panel: CI contributions in our quasi-diabatic representation (not shown for $\phi=0$ because they would involve ill-defined compensations between spurious numerical infinities). Spline  interpolations have been used to smooth the curves. Bottom panel: CSF contributions (equal to 0 for both $\phi=0$ and $\phi=90$ degree).
    }  \label{fig:nac_theta_singleplot}
\end{figure}

Both maps are plotted in Fig. \ref{fig:descriptor_d_and_r}.
We observe that $d(\bm{q})$ varies little and smoothly between about 0.09 (``pole'' origin) and 0.14, which is small by about an order of magnitude compared to $\sqrt{2}$.
Then, we observe that $r(\bm{q})$ also varies little and smoothly between a strict 0 (``pole'' origin) and about 0.11, which remains small by more than about an order of magnitude compared to $2\sqrt{2}$.
The deviation from optimality, given by $r(\bm{q})$, follows symmetry considerations: it is strictly zero for $\phi=0^\circ$ or $90^\circ$ ($\mathrm{C_s}$ geometries where the adiabatic and diabatic representations coincide locally).
Interestingly enough, its smallest finite values occur within the most critical nonadiabatic region (near the conical intersection: look at $\phi=2.5^\circ$), which is a favorable argument for considering that our quasi-diabatic representation -- obtained ``for free'' -- is almost optimally diabatic where it has to be so.
However, the fact that $d(\bm{q})$ is never zero reflects an incompressible measure of the intrinsic validity of a two-state block-Born-Oppenheimer approximation for this particular molecule.

Once again, let us recall that in the present work, we did not enforce optimality as regards diabaticity, except for using diabatized orbitals.
The above observations simply tell us that the overall algorithm naturally brings a representation that is almost optimally diabatic ``for free'' and with consistent signs.
This is a strong incentive for making use of such a property if we want QC techniques to be used ``as lightly as possible'' (avoiding unnecessary transformations or constraints, likely to induce extra noise), especially in the context of direct quantum dynamics, as already alluded to.

Finally, let us examine the behavior of the various NAC contributions.
Let us recall that the use of diabatic orbitals (see Appendix~\ref{app:diaborbs}) as defined by Werner \emph{et al.}~\cite{wer88:3139,sim99:4523} is specifically meant to minimize the nonremovable CSF contribution to the NACs in a balanced manner at all geometries.
The maps of $\bm{F}_{01}^\mathrm{CSF}(\bm{q}) = \bm{f}_{01}^\mathrm{CSF}(\bm{q})$ as derivatives with respect to $\phi$ and $\theta$ are plotted in Figs. \ref{fig:nac_phi_singleplot} and \ref{fig:nac_theta_singleplot}, bottom panels, respectively.
Such magnitudes are dimensionless (or more precisely given in $\rm rad^{-1}$).
As such, they are to be compared to a typical value of $\pm 1~\rm rad^{-1}$ as regards the evaluation of their ``smallness''.
As expected, both CSF-NAC components behave smoothly and can be considered as being negligible everywhere (their typical order of magnitude is about $\pm 0.01~\rm rad^{-1}$).

In the same figures, we see that the residual (quasi-diabatic) CI-NACs remain limited to values that do not exceed a few $\pm 0.01~\rm rad^{-1}$; see middle panels of Figs. \ref{fig:nac_phi_singleplot} and \ref{fig:nac_theta_singleplot}.
The deviation from diabatic optimality of our quasi-diabatic CI-NACs thus does not exceed the order of magnitude of the irremovable CSF-NACs, so that we can confirm that our quasi-diabatic representation is diabatic enough from a numerical point of view.
Finally, the adiabatic CI-NACs tend to get values that are about three orders of magnitude larger when they remain definite (about $\pm 10~\rm rad^{-1}$), which sets a scale for not neglecting NACs around a conical intersection; see top panels of Figs. \ref{fig:nac_phi_singleplot} and \ref{fig:nac_theta_singleplot}.
As expected, they diverge at the conical intersection, which numerically brings them to ill-determined values of about $\pm 10^3-10^4~\rm rad^{-1}$ (far beyond the scales used in Figs. \ref{fig:nac_phi_singleplot} and \ref{fig:nac_theta_singleplot}, top panels).


\section{Conclusions and outlook}\label{sec:conclu}

In conclusion, the present study establishes that the state-average orbital-optimized variational quantum eigensolver (SA-OO-VQE) naturally produces an \emph{ab initio} quasi-diabatic representation through its inherent least-transformed block-diagonalization property. 
This capability, previously hypothesized in Ref.~\cite{yal21:024004}, has been rigorously validated here using diabaticity descriptors and demonstrated on the formaldimine molecule, which features a well-characterized conical intersection between its ground and first-excited singlet states.
These results underscore the practical utility and potential of SA-OO-VQE for addressing molecular systems that undergo complex nonadiabatic phenomena.
Additional analytical derivations meant to further assess the detailed properties of this least-transformed block-diagonalization are left for future work.

The diabatic states generated by this method are expected to enable quantum dynamics simulations of nuclear wavepacket propagation on smooth diabatic potential energy surfaces. 
These applications include the direct-dynamics variational multiconfiguration Gaussian (DD-vMCG) approach with on-the-fly evaluations of the electronic Hamiltonian matrix or the multiconfigurational time-dependent Hartree (MCTDH) method with parameterized vibronic-coupling Hamiltonian models. 
For such studies, obtaining the gradient of the off-diagonal Hamiltonian element 
$H_{\rm AB}$ will be essential; it will be calculated in analogy to the analytical gradients exposed in Ref.~\cite{yal22:776}.

Extending this approach to more challenging systems, in particular beyond two electronic states, is a critical future direction as many ultrafast photochemical and photophysical processes involve avoided crossings and conical intersections between multiple states. 
Finally, if the least-transformed block-diagonalization approach does not consistently yield optimal diabaticity, incorporating nonadiabatic couplings directly into the SA-OO-VQE cost function could be a solution.
To achieve this efficiently, one could compute the couplings analytically without relying on their values in the adiabatic representation, in contrast to this work [Eq.~(\ref{eq:NAC_dia_adia})].
A simpler solution would be to incorporate the residual diabatic descriptor $r({\bmq})$ into the cost function.
Addressing such challenges represents an exciting avenue for further research.

\section*{Acknowledgments}

This work was publicly funded through ANR (the French National Research Agency) under
the ``Investissements d'avenir'' program with the reference ANR-16-IDEX-0006. This work was supported by the Ministry of Education, Youth and Sports of the Czech Republic through the e-INFRA CZ (ID:90254). 

\section*{Code and data availability}

The SA-OO-VQE solver code, interfaced with \textsc{Psi4} \cite{psi4} for the classical calculation of the electronic integrals, is published at the repository \url{https://gitlab.com/MartinBeseda/sa-oo-vqe-qiskit.git}. 
This work was using version 1.2.0, which is also published under Zenodo DOI \texttt{10.5281/zenodo.14596890}. 
The Zenodo repository for v1.2.0 also contains the reference data to this paper and the scripts, which can be used to reproduce the results in the \texttt{sa-oo-vqe-qiskit/example/diabatization} folder.
In addition to SA-OO-VQE QC emulations, classical-computing implementation of SA-MCSCF was used herein for geometry optimizations at the SA(2)-CAS(4,3)SCF/cc-pVDZ level of theory with equal weights on $\rm S_0$ and $\rm S_1$ with \textsc{Gaussian16} \cite{g16}.
Note that GUCCSD with equal-weighted SA-OO-VQE is equivalent here to SA(2)-CAS(4,3)SCF since it is capable of generating the six singlet CSFs. 
We checked that the energies calculated with both SA-OO-VQE and SA-CASSCF matched within about $10^{-7}$ hartree at all geometries considered in the present work.

\appendix

\section{Geometries of the formaldimine molecule}\label{app:coord}

The Cartesian coordinates of the atoms of the formaldimine molecule, explored herein around the `reference conical intersection', are given in Table~\ref{tab:coord},
where
\begin{eqnarray}
X &=& R \sin\theta \cos\phi \nonumber \quad, \\
Y &=& R \sin\theta\sin\phi \nonumber \quad, \\
Z &=& - R \cos\theta \quad. \nonumber
\end{eqnarray}
\begin{table}[!ht]
\caption{Cartesian coordinates (in \AA) of the formaldimine molecule in this work. }
\begin{tabular}{cccc}
Atom & $x$ & $y$ & $z$ \\
N   &               0.00000000  &  0.00000000  &  0.00000000\\
C   &               0.00000000  &  0.00000000  & -1.41205200\\
H   &               0.00000000  & -0.94421526  & -1.93740123\\
H   &               0.00000000  &  0.94421526  & -1.93740123\\
H   &               $X$         &   $Y$        &  $Z$
\end{tabular}
\label{tab:coord}
\end{table}

At the `reference conical intersection' (partially optimised as a minimum-energy conical intersection under the constraint that 
the $\mathrm{H_2CN}$-fragment remains planar), we have $R = 1.017251$ \AA, $\theta=109.5902^\circ$, and $\phi=90^\circ$.
It was obtained at the SA(2)-CAS(4,3)SCF/cc-pVDZ level of theory with equal weights on $\rm S_0$ and $\rm S_1$, with \textsc{Gaussian16} \cite{g16}.
The energies are: 
$E_0 = -93.95339~\mathrm{hartree}$ and $
E_1 = -93.95338~\mathrm{hartree}
$ (energy difference: $0.0002~\mathrm{eV}$).

\section{Diabatic orbitals}\label{app:diaborbs}

As described in Sec.~\ref{sec:modsys}, we explored the surroundings of the conical intersection between the
ground and first-excited state of the formaldimine molecule by considering variations of two spherical angles, ($\theta,\phi$), and we used
reference orbitals as starting orbitals for
each different molecular geometry.
In this Appendix, we describe how the diabatic orbitals are obtained. 
We use $\mu$ and $\nu$ to denote atomic orbitals and
$p$ and $q$ for active molecular orbitals.
The reference geometry has been chosen to be the high-symmetry origin, $\mathrm{C_{2v}}$ planar geometry, as described in Sec.~\ref{sec:modsys}, and will be denoted by $\bmq^0$.

The active canonical molecular orbitals chosen as a reference for further diabatization will be denoted $\ket{\psi_{p}; \bmq^0}=\ket{\psi^0_{p}}$ and the atomic orbitals $\ket{\chi_{\mu}; \bmq^0}=\ket{\chi^0_{\mu}}$. The former are obtained after performing the SA-OO-VQE algorithm, \emph{i.e.}, they are the native state-average optimized orbitals at geometry $\bmq^0$.
They are
written as linear combinations of atomic orbitals (LCAO) as follows,
\begin{equation}
    \ket{\psi^0_{p}} = \sum_{\mu} x^0_{\mu p} \ket{\chi^0_{\mu}} \quad.
\end{equation}

Now, starting from
the canonical (Hartree--Fock) state-average optimized orbitals at geometry $\bmq$,
\begin{equation}
    \ket{\psi_{p};\bmq} = \sum_{\mu} x_{\mu p}(\bmq) \ket{\chi_{\mu}; \bmq} \quad,
\end{equation}
we want to obtain the diabatic orbitals at $\bmq$ that are maximally overlapping with the reference orbitals at $\bmq^0$.
Let us write the overlap matrix $\mathbf{c}(\bmq)$,
\begin{equation}
    c_{pq}(\bmq) = \psh{\psi^0_p}{\psi_q;\bmq} = \sum_{\mu,\nu} x^{0*}_{\mu p} s^{\rm AO}_{\mu\nu} (\bmq) x_{\nu q}(\bmq) \quad,
\end{equation}
where
\begin{equation}
    s^{\rm AO}_{\mu \nu}(\bmq) = \psh{\chi^0_{\mu}}{\chi_\nu ; \bmq} \quad,
\end{equation}
is the overlap matrix between atomic orbitals at the reference ($\bmq^0$) and current geometries ($\bmq$).
Following the exact same strategy as for the quasi-diabatic many-body states,
we employ a L\"owdin Hermitian orthonormalization, as detailed in Eq.~(\ref{eq:T_Lowdin}) and in Appendix~\ref{app:lowdin}.
In order to discriminate the one-body basis used in this Appendix from the many-body basis invoked elsewhere, we use lowercase matrix symbols.
The diabatic orbitals at $\bmq$, 
$\lbrace \ket{\phi_p;\bmq} \rbrace$, 
are then obtained as
follows,
\begin{eqnarray}
\mathbf{o}(\bmq)
= \mathbf{c}(\bmq) \mathbf{t}^\dagger(\bmq) \quad,
\end{eqnarray}
where 
$o_{pq}(\bmq) = \psh{\psi^0_p}{\phi_q; \bmq}$
and
$\mathbf{t}(\bmq)$ is the unitary transformation matrix that reads
\begin{eqnarray}
\mathbf{t}(\bmq) = \mathbf{c}^\dagger(\bmq) (\mathbf{c}(\bmq)\mathbf{c}^\dagger(\bmq))^{-1/2} \quad,
\end{eqnarray}
with
$t_{pq}(\bmq) = \psh{\phi_p; \bmq}{\psi_q;\bmq}$.
In complete analogy with the many-body problem, the matrix $\mathbf{o}(\bmq)$ will be as close as possible to the identity matrix.

While this step may seem innocent at first sight, it actually is central as regards the whole procedure because it provides a well-defined reference with sign consistency set at the one-body level rather than at the many-body one. Similar considerations are to be found in Ref. \cite{mer23:1827}.

\section{Various contributions to the nonadiabatic couplings in different basis sets}\label{app:NAC}

Let us consider three types of orthonormal many-body basis sets, expected to be numerically complete with respect to a CI-type variational Hilbert manifold of finite size $N>0$
(typically, the number of CSFs), containing both the model and target Hilbert subspaces (of size $2\leq N$ in the main text). 
Herein, their parametric $\bm{q}$-dependence will be made implicit for notational simplicity.

The ``reference'' 
basis set, $\{\ket{\Phi^0_J}\}_{J=0}^{N-1}$, is assumed to be smooth and regular with respect to $\bm{q}$-variations (it is often taken as real-valued in practice but this is not a formal requirement); 
furthermore, we suppose that it is as diabatic as possible (see Appendix~\ref{app:lowdin}).
For us, this will be the CSF-basis set built on MOs that 
we want to be as diabatic as possible, determined by the optimal $\bmkappa$\martinfix{-}{ } parameters and some unitary post-processing orbital-diabatization procedure (see Appendix~\ref{app:diaborbs}).

The ``intermediate'' 
(expected to be quasi-diabatic later on) basis set, $\{\ket{\Phi_L}\}_{L=0}^{N-1}$, is also assumed to be smooth and regular with respect to $\bm{q}$-variations and be such that it makes the CI Hamiltonian matrix block-diagonal with respect to the target.
In practice, this is the variational solution to the SA-OO-VQE objective, determined by the optimal $\bmtheta$-parameters, knowing the optimal $\bmkappa$-parameters. 
As mentioned in the main text, we shall focus on $\{\ket{\Phi_0},\ket{\Phi_1}\}\equiv\{\ket{\Phi_\rmA},\ket{\Phi_\rmB}\}$ (the target) as being as close as possible to $\{\ket{\Phi^0_0},\ket{\Phi^0_1}\}\equiv\{\ket{\Phi^0_\rmA},\ket{\Phi^0_\rmB}\}$ (the model, meant to be a ``good guess'' of the target from a ``least-transformed'' criterion).

The ``adiabatic'' basis set, $\{\ket{\Psi_\alpha}\}_{\alpha=0}^{N-1}$, is defined through a post-variational principle from the previous one 
within the target subspace, spanned by either $\{\ket{\Psi_0},\ket{\Psi_1}\}$ or $\{\ket{\Phi_\rmA},\ket{\Phi_\rmB}\}$, both related 
via a restricted unitary transformation (here, represented by a $(2 \times 2)$-change-of-basis-matrix, but this can be generalized to larger subspaces).
Hence, the first two adiabatic states are specifically defined so as to solve the final partial eigenproblem within the target subspace, while 
the remaining states can be left unchanged.
In practice, the first two adiabatic 
states, $\{\ket{\Psi_0},\ket{\Psi_1}\}$, are determined from $\{\ket{\Phi_\rmA},\ket{\Phi_\rmB}\}$ by a rotation through the terminal post-variational $\varphi$-parameter.

Once again, let 
us stress that the specificity of our SA-OO-VQE approach, compared to standard SA-MCSCF implementations, is that we have 
the liberty to stop the actual variational procedure at the block-diagonalization stage (true objective: 
optimal $\bmtheta$-parameters, knowing the optimal $\bmkappa$-parameters), before the final post-variational 
rotation (practical objective: terminal $\varphi$-parameter).

Hence, we have
\begin{subequations}
\begin{align}
& \ket{\Psi_\alpha} = \sum_J C_{J\alpha} \ket{\Phi^0_J} = \sum_L T_{L\alpha} \ket{\Phi_L} \quad, \\
& \ket{\Phi_L} = \sum_J O_{JL} \ket{\Phi^0_J} \quad,
\end{align}
\end{subequations}
where
\begin{subequations}
\begin{align}
& C_{J\alpha} = \braket{\Phi^0_J}{\Psi_\alpha} \quad, \\
& O_{JL} = \braket{\Phi^0_J}{\Phi_L} \quad, \\
& T_{L\alpha} = \braket{\Phi_L}{\Psi_\alpha} \quad,
\end{align}
\end{subequations}
and
\begin{equation}
\bold{C} = \bold{O} \bold{T} \quad.
\end{equation}
For the moment, 
the three $(N \times N)$-matrices are supposed to be determined and unitary [in practice, we shall only determine a 
$(2 \times 2)$-block for $\bold{T}$ and $(N \times 2)$-blocks for $\bold{C}$ and $\bold{O}$]. 

Now, the corresponding matrix elements of the 
first-order NACs are defined as
\begin{subequations}
\begin{align}
& \bold{F}_{\alpha\beta} = \braket{\Psi_\alpha}{\bm\nabla\Psi_\beta} \quad, \\
& \bold{f}_{LM} = \braket{\Phi_L}{\bm\nabla\Phi_M} \quad, \\
& \bold{f}^0_{JK} = \braket{\Phi^0_J}{\bm\nabla\Phi^0_K} \quad, 
\end{align}
\end{subequations}
and related via
\begin{subequations}
\begin{align}
& \bold{F} = \bold{C}^\dag\bm\nabla\bold{C} + \bold{C}^\dag \bold{f}^0 \bold{C} = \bold{T}^\dag\bm\nabla\bold{T} + \bold{T}^\dag \bold{f} \bold{T} \quad, \\ 
& \bold{f} = \bold{O}^\dag\bm\nabla\bold{O} + \bold{O}^\dag \bold{f}^0 \bold{O} \quad, 
\end{align}
\end{subequations}
such that
\begin{subequations}
\begin{align}
& \bold{F} = \bold{F}^\mathrm{CI} + \bold{F}^\mathrm{CSF} \quad, \\ 
& \bold{f} = \bold{f}^\mathrm{CI} + \bold{f}^\mathrm{CSF} \quad, 
\end{align}
\end{subequations}
with
\begin{subequations}
\begin{align}
& \bold{F}^\mathrm{CI} = \bold{C}^\dag\bm\nabla\bold{C}\quad, & \bold{F}^\mathrm{CSF} = \bold{C}^\dag\bold{f}^0\bold{C} \quad, \\
& \bold{f}^\mathrm{CI} = \bold{O}^\dag\bm\nabla\bold{O}\quad, & \bold{f}^\mathrm{CSF} = \bold{O}^\dag\bold{f}^0\bold{O} \quad.
\end{align}
\end{subequations}
Then, defining
\begin{equation}
\bold{A} = \bold{T}^\dag\bm\nabla\bold{T} \quad,
\end{equation}
and owing to 
\begin{equation}
\bold{C}^\dag\bm\nabla\bold{C} = \bold{T}^\dag \bold{O}^\dag\bm\nabla\bold{O} \bold{T} + \bold{T}^\dag \bold{O}^\dag\bold{O} \bm\nabla\bold{T} \quad,
\end{equation}
we get
\begin{subequations}
\begin{align}
& \bold{F}^\mathrm{CI} = \bold{T}^\dag \bold{f}^\mathrm{CI} \bold{T} + \bold{A} \quad, \\
& \bold{F}^\mathrm{CSF} = \bold{T}^\dag \bold{f}^\mathrm{CSF} \bold{T} \quad.
\end{align}
\end{subequations}

Let us remark that all NAC-type square matrices defined above are skew-Hermitian. 
This is proved by differentiating the conditions of orthonormality of 
the basis sets and unitarity of the coefficient matrices. For example,
\begin{subequations}
\begin{align}
& \braket{\Phi^0_J}{\Phi^0_K} = \delta_{JK} \quad, \\
& \braket{\bm\nabla\Phi^0_J}{\Phi^0_K} + \braket{\Phi^0_J}{\bm\nabla\Phi^0_K} = \bm0 \quad, \\
& \braket{\Phi^0_K}{\bm\nabla\Phi^0_J}^* = -\braket{\Phi^0_J}{\bm\nabla\Phi^0_K} \quad,
\end{align}
\end{subequations}
and
\begin{subequations}
\begin{align}
& \bold{C}^\dag\bold{C} = \bm1 \quad, \\
& \bm\nabla\bold{C}^\dag\bold{C} + \bold{C}^\dag\bm\nabla\bold{C} = \bm0 \quad, \\
& (\bold{C}^\dag\bm\nabla\bold{C})^\dag = -\bold{C}^\dag\bm\nabla\bold{C} \quad,
\end{align}
\end{subequations}
and similarly for all the others.

Now, let us turn to the electronic Hamiltonian matrix representations,
\begin{subequations}
\begin{align}
& H^0_{JK} =\mel{\Phi^0_J}{\hat{H}^{\rm{el}}}{\Phi^0_K}\quad, \\
& H_{LM} =\mel{\Phi_L}{\hat{H}^{\rm{el}}}{\Phi_M} \quad, \\
& E_{\alpha\beta} =\mel{\Psi_\alpha}{\hat{H}^{\rm{el}}}{\Psi_\beta} \quad,
\end{align}
\end{subequations}
and investigate first-order Hellmann--Feynman-type formulae. The three Hamiltonian matrices are related via unitary similarity transformations, 
\begin{subequations}
\begin{align}
& \bold{E} = \bold{C}^\dag \bold{H}^0 \bold{C} = \bold{T}^\dag\bold{O}^\dag \bold{H}^0 \bold{O}\bold{T} = \bold{T}^\dag \bold{H} \bold{T} \quad, \\
& \bold{H} = \bold{O}^\dag \bold{H}^0 \bold{O} \quad.
\end{align}
\end{subequations}

The Hellmann--Feynman machinery starts with differentiating the overlap matrices, 
which in practice implies that all sorts of NAC matrices are skew-Hermitian (see above). 
The second step consists of differentiating the Hamiltonian matrices, which yields, upon inserting 
the resolution of the identity in matrix form (\emph{e.g.}, $\bold{C}\bold{C}^\dag=\bm1$), 
\begin{subequations}
\begin{align}
& \bm\nabla\bold{E} = \bold{E}\bold{F}^\mathrm{CI} - \bold{F}^\mathrm{CI}\bold{E} + \bold{C}^\dag \bm\nabla\bold{H}^0 \bold{C} \quad, \\
& \bm\nabla\bold{E} = \bold{E}\bold{A} - \bold{A}\bold{E} + \bold{T}^\dag \bm\nabla\bold{H} \bold{T} \quad, \\
& \bm\nabla\bold{H} = \bold{H}\bold{f}^\mathrm{CI} - \bold{f}^\mathrm{CI}\bold{H} + \bold{O}^\dag \bm\nabla\bold{H}^0 \bold{O} \quad.
\end{align}
\end{subequations}
It must be understood that such Hellmann--Feynman-type relations (within the CI variational manifold of size $N$ spanned by the reference CSFs) are exact so far.
In contrast, starting from the operator form (within the infinite Hilbert space of kets) yields extra terms called Pulay forces that are a manifestation of the fact that the Schrödinger problem is not exactly solved when using a variational CI approach.
In simple words, the key is that the gradient of an orbital at some geometry has no reason to belong to the finite manifold of orbitals at this geometry.

Until now, we have not specified any particular type of subspace partitioning. 
The previous relations only reflect that we have been considering three different basis sets and unitary similarity transforms of Hamiltonian matrices. 
In what follows, we specifically assume that the main purpose of $\bold{O}$ is to achieve a block-diagonalization of the Hamiltonian matrix while 
$\bold{T}$ only terminates its partial diagonalization within the target subspace. 
Note that square submatrix restrictions within it will be denoted with a ``breve''
symbol. 

First, we get
\begin{subequations}
\begin{align}
& \bold{\breve{A}} = \bold{\breve{T}}^\dag\bm\nabla\bold{\breve{T}} \quad, \\
& \bold{\breve{F}}^\mathrm{CI} = \bold{\breve{T}}^\dag \bold{\breve{f}}^\mathrm{CI} \bold{\breve{T}} + \bold{\breve{A}} \quad, \\
& \bold{\breve{F}}^\mathrm{CSF} = \bold{\breve{T}}^\dag \bold{\breve{f}}^\mathrm{CSF} \bold{\breve{T}} \quad,
\end{align}
\end{subequations}
with
\begin{equation}
\bold{T} = 
\begin{pmatrix}
\bold{\breve{T}} & \bm0 \\
\bm0 & \bm1
\end{pmatrix} \quad, \quad
\bold{A} = 
\begin{pmatrix}
\bold{\breve{A}} & \bm0 \\
\bm0 & \bm0
\end{pmatrix} \quad,
\end{equation}
where $\bold{\breve{T}}$ is unitary and $\bold{\breve{A}}$ is skew-Hermitian.
Reciprocally, we have 
\begin{subequations}
\begin{align}
& \bold{\breve{T}}\bold{\breve{A}}\bold{\breve{T}}^\dag = - \bold{\breve{T}}\bm\nabla\bold{\breve{T}}^\dag \quad, \\
& \bold{\breve{f}}^\mathrm{CI} = \bold{\breve{T}}(\bold{\breve{F}}^\mathrm{CI} - \bold{\breve{A}})\bold{\breve{T}}^\dag \quad, \\
& \bold{\breve{f}}^\mathrm{CSF} = \bold{\breve{T}}\bold{\breve{F}}^\mathrm{CSF}\bold{\breve{T}}^\dag \quad.
\end{align}
\end{subequations}
These relations are central to the present work.
They provide an exact way for evaluating the ``residual'' NACs in the intermediate basis from the analytic ones obtained in the adiabatic basis.
They have been made more transparent in the main text when addressing a 
two-state problem specifically. 

We can continue the Hellmann--Feynman machinery. The Hamiltonian matrices and their gradients have the following block structures, 
\begin{subequations}
\begin{align}
&\bold{H} = 
\begin{pmatrix}
\bold{\breve{H}} & \bm0 \\
\bm0 & \bold{\bar{H}}
\end{pmatrix} \quad, \quad
\bold{E} = 
\begin{pmatrix}
\bold{\breve{E}} & \bm0 \\
\bm0 & \bold{\bar{H}}
\end{pmatrix} \quad, \\
&\bm\nabla\bold{H} = 
\begin{pmatrix}
\bm\nabla\bold{\breve{H}} & \bm0 \\
\bm0 & \bm\nabla\bold{\bar{H}}
\end{pmatrix} \quad, \quad
\bm\nabla\bold{E} = 
\begin{pmatrix}
\bm\nabla\bold{\breve{E}} & \bm0 \\
\bm0 & \bm\nabla\bold{\bar{H}}
\end{pmatrix} \quad.
\end{align}
\end{subequations}
Note that the complementary submatrices, $\bold{\bar{H}}$ and $\bm\nabla\bold{\bar{H}}$, 
are not to be determined. 

The Hamiltonian target submatrices satisfy
\begin{equation}
\bold{\breve{E}} = \bold{\breve{T}}^\dag \bold{\breve{H}} \bold{\breve{T}} \quad, \quad \bm1 = \bold{\breve{T}}^\dag \bold{\breve{T}} \quad,
\end{equation}
hence,
\begin{equation}
\bold{\breve{H}} \bold{\breve{T}} = \bold{\breve{T}} \bold{\breve{E}} \quad.
\end{equation}
If $\bold{\breve{E}}$ 
is further assumed to be a diagonal submatrix, the latter relation is a linear variational (secular) and partial 
eigenproblem, which is exact with respect to $\bold{\breve{H}}$ but not $\hat{H}^{\rm{el}}$.

Now, we can examine how the Hellmann--Feynman-type relations manifest themselves within the target subspace. 
Let us denote
\begin{equation}
\bold{G} = \bold{O}^\dag \bm\nabla\bold{H}^0 \bold{O} \quad,
\end{equation}
and $\bold{\breve{G}}$ its target restriction. Their matrix elements are to be understood as commensurate to energy gradient vectors. We thus get 
\begin{subequations}
\begin{align}
& \bm\nabla\bold{\breve{E}} = \bold{\breve{E}}\bold{\breve{F}}^\mathrm{CI} - \bold{\breve{F}}^\mathrm{CI}\bold{\breve{E}} + \bold{\breve{T}}^\dag\bold{\breve{G}}\bold{\breve{T}} \quad, \\
& \bm\nabla\bold{\breve{E}} = \bold{\breve{E}}\bold{\breve{A}} - \bold{\breve{A}}\bold{\breve{E}} + \bold{\breve{T}}^\dag \bm\nabla\bold{\breve{H}} \bold{\breve{T}} \quad, \\
& \bm\nabla\bold{\breve{H}} = \bold{\breve{H}}\bold{\breve{f}}^\mathrm{CI} - \bold{\breve{f}}^\mathrm{CI}\bold{\breve{H}} + \bold{\breve{G}} \quad.
\end{align}
\end{subequations}

Because $\bold{\breve{E}}$ -- and \emph{a fortiori} $\bm\nabla\bold{\breve{E}}$ -- are assumed to be diagonal submatrices, we obtain the Hellmann--Feynman formulae for the gradients and CI-NACs as follows (assuming, here, real-valued states). 
First, 
\begin{equation}
\bm\nabla\mathcal{E}_\alpha \delta_{\alpha\beta} = (\mathcal{E}_\alpha - \mathcal{E}_\beta) \bm{F}_{\alpha\beta}^\mathrm{CI}  + [\bold{\breve{T}}^\dag \bold{\breve{G}} \bold{\breve{T}}]_{\alpha\beta} \quad,
\end{equation}
\emph{i.e.},
\begin{subequations}
\begin{align}
& \bm\nabla\mathcal{E}_\alpha = [\bold{\breve{T}}^\dag \bold{\breve{G}} \bold{\breve{T}}]_{\alpha\alpha} \quad, \\
& \bm{F}_{\alpha\beta}^\mathrm{CI} = (1-\delta_{\alpha\beta})\frac{[\bold{\breve{T}}^\dag \bold{\breve{G}} \bold{\breve{T}}]_{\alpha\beta}}{\mathcal{E}_\beta - \mathcal{E}_\alpha} \quad.
\end{align}
\end{subequations}
Second, 
\begin{equation}
\bm\nabla\mathcal{E}_\alpha \delta_{\alpha\beta} = (\mathcal{E}_\alpha - \mathcal{E}_\beta) \bm{A}_{\alpha\beta}  + [\bold{\breve{T}}^\dag \bm\nabla\bold{\breve{H}} \bold{\breve{T}}]_{\alpha\beta} \quad,
\end{equation}
\emph{i.e.},
\begin{subequations}
\begin{align}
& \bm\nabla\mathcal{E}_\alpha = [\bold{\breve{T}}^\dag \bm\nabla\bold{\breve{H}} \bold{\breve{T}}]_{\alpha\alpha} \quad, \\
& \bm{A}_{\alpha\beta} = (1-\delta_{\alpha\beta})\frac{[\bold{\breve{T}}^\dag \bm\nabla\bold{\breve{H}} \bold{\breve{T}}]_{\alpha\beta}}{\mathcal{E}_\beta - \mathcal{E}_\alpha} \quad.
\end{align}
\end{subequations}

In addition, we have
\begin{equation}
[\bold{\breve{T}}^\dag (\bm\nabla\bold{\breve{H}}-\bold{\breve{G}})\bold{\breve{T}}]_{\alpha\beta} = (\mathcal{E}_\alpha - \mathcal{E}_\beta) (\bm{F}_{\alpha\beta}^\mathrm{CI} - \bm{A}_{\alpha\beta}) \quad,
\end{equation}
\emph{i.e.},
\begin{subequations}
\begin{align}
& [\bold{\breve{T}}^\dag \bm\nabla\bold{\breve{H}} \bold{\breve{T}}]_{\alpha\alpha} = [\bold{\breve{T}}^\dag \bold{\breve{G}} \bold{\breve{T}}]_{\alpha\alpha} \quad, \\
& \bm{F}_{\alpha\beta}^\mathrm{CI} - \bm{A}_{\alpha\beta} = (1-\delta_{\alpha\beta}) \frac{[\bold{\breve{T}}^\dag (\bold{\breve{G}}-\bm\nabla\bold{\breve{H}})\bold{\breve{T}}]_{\alpha\beta}}{\mathcal{E}_\beta - \mathcal{E}_\alpha} \quad,
\end{align}
\end{subequations}
from which we can assemble 
\begin{equation}
\bm{f}_{\alpha\beta}^\mathrm{CI} = [\bold{\breve{T}}(\bold{\breve{F}}^\mathrm{CI} - \bold{\breve{A}})\bold{\breve{T}}^\dag]_{\alpha\beta} \quad.
\end{equation}
This is illustrated in the main text in the two-state case.

Note that we have assumed real-valued states here. 
As a consequence, the diagonal entries of the skew-Hermitian matrices vanish when they are finite (hence the $(1-\delta_{\alpha\beta})$-factor). 
Such relations imply some strong regular differentiation properties, whereby issues related to Berry/Longuet-Higgins geometric/topological phase -- which are beyond the scope of the present work -- are omitted \cite{bae06}. 
In other words, the NAC matrices are assumed to have, globally, zero diagonal entries, but potentially infinite, locally, if a phase discontinuity is encountered. 
In fact, such local singularities are both met by $\bold{\breve{F}}^\mathrm{CI}$ and $\bold{\breve{A}}$ but should fully compensate in $\bold{\breve{f}}^\mathrm{CI}=\bold{\breve{T}}(\bold{\breve{F}}^\mathrm{CI}-\bold{\breve{A}})\bold{\breve{T}}^\dag$, which is meant to be smooth and regular everywhere -- as detailed in Ref. \cite{bae06} -- as long as the target subspace remains nonsingular (no energy crossing with its complement and a valid block-Born-Oppenheimer approximation).

Now, the previous discussion only addressed the CI contribution to the NACs. 
Yet, if we want to produce a representation that is as diabatic as possible, 
we have to address the CSF contribution to the NACs too, which is a nonremovable term related to Pulay forces and CP-MCSCF contributions.
It is obtained via SA-MCSCF-type analytic derivative techniques \cite{len92:1,yal22:776}.
Owing to gauge freedom, it can be minimized upon rotating the MO basis set such that it varies as little as possible with the nuclear coordinates, as proposed by Werner \emph{et al.} who 
introduced the concept of diabatic orbitals and implemented it in the MOLPRO quantum-chemistry software \cite{wer88:3139,sim99:4523}.
Note that this problem was re-explored recently within the context of the TSH algorithm available in the MOLCAS quantum-chemistry software \cite{mer23:1827}. 
As shown in Appendix~\ref{app:diaborbs}, this boils down to a Hermitian Löwdin/unitary Procrustes problem, based on an SVD for maximal LCAO-coefficient overlap matrix with respect to a reference geometry. 
This procedure for the one-body problem is very similar in spirit to what is shown below for the many-body one. 

\section{Least-transformed quasi-diabatic block-diagonalization via Löwdin Hermitian orthonormalization, polar and square singular-value decompositions, or unitary Procrustes analysis}\label{app:lowdin}

The least-transformed block-diagonalization problem and its relation with quasi-diabatization has been known in the literature under various equivalent formulations. 
The objective of the present Appendix is not to be exhaustive but to clarify and facilitate the identification of essential connections among them. 

Most quantities involved below have been defined in the above as regards the CI many-body and orbital one-body problems involving CSFs up to some gauge freedom.
What follows will sometimes be exemplified with a $2:(N-2)$-partition for simplicity (with extra details to be found in the main text), but full generalization is straightforward.

\subsection{From the target to the model}

Let us first consider the target subspace as a given and assume the well-posed existence of a model subspace. Within a CI many-body context, the two nonsingular (invertible) target-to-model submatrices of interest (aimed at partial diagonalization and block-diagonalization) are
\begin{subequations}
\begin{align}
& \bold{\breve{C}} = 
\begin{pmatrix}
C_{\rmA0} & C_{\rmA1} \\
C_{\rmB0} & C_{\rmB1}
\end{pmatrix} =
\begin{pmatrix}
\braket{\Phi^0_\rmA}{\Psi_0} & \braket{\Phi^0_\rmA}{\Psi_1} \\
\braket{\Phi^0_\rmB}{\Psi_0} & \braket{\Phi^0_\rmB}{\Psi_1}
\end{pmatrix} \quad, \\
& \bold{\breve{O}} = 
\begin{pmatrix}
O_{\rmA\rmA} & O_{\rmA\rmB} \\
O_{\rmB\rmA} & O_{\rmB\rmB}
\end{pmatrix} =
\begin{pmatrix}
\braket{\Phi^0_\rmA}{\Phi_\rmA} & \braket{\Phi^0_\rmA}{\Phi_\rmB} \\
\braket{\Phi^0_\rmB}{\Phi_\rmA} & \braket{\Phi^0_\rmB}{\Phi_\rmB}
\end{pmatrix} \quad.
\end{align}
\end{subequations}
They are not unitary but expected to be close to it (in particular, $\bold{\breve{O}}$ is expected to be close to the identity; see later on).
Both matrices are related via a unitary matrix, 
\begin{equation}
\bold{\breve{T}} = 
\begin{pmatrix}
T_{\rmA0} & T_{\rmA1} \\
T_{\rmB0} & T_{\rmB1}
\end{pmatrix} =
\begin{pmatrix}
\braket{\Phi_\rmA}{\Psi_0} & \braket{\Phi_\rmA}{\Psi_1} \\
\braket{\Phi_\rmB}{\Psi_0} & \braket{\Phi_\rmB}{\Psi_1}
\end{pmatrix} \quad,
\end{equation}
such that, as already mentioned, $\bold{\breve{C}}=\bold{\breve{O}}\bold{\breve{T}}$.

From there, we can invoke the projector onto the model subspace,
\begin{equation}
\hat{P}^0 = \ket{\Phi^0_\rmA} \bra{\Phi^0_\rmA} +\ket{\Phi^0_\rmB} \bra{\Phi^0_\rmB} \quad,
\end{equation}
such that
\begin{subequations}
\begin{align}
& \hat{P}^0 \ket{\Psi_0} = C_{\rmA0} \ket{\Phi^0_\rmA} + C_{\rmB0} \ket{\Phi^0_\rmB} \quad, \\
& \hat{P}^0 \ket{\Psi_1} = C_{\rmA1} \ket{\Phi^0_\rmA} + C_{\rmB1} \ket{\Phi^0_\rmB} \quad,
\end{align}
\end{subequations}
and
\begin{subequations}
\begin{align}
& \hat{P}^0 \ket{\Phi_\rmA} = O_{\rmA\rmA} \ket{\Phi^0_\rmA} + O_{\rmB\rmA} \ket{\Phi^0_\rmB} \quad, \\
& \hat{P}^0 \ket{\Phi_\rmB} = O_{\rmA\rmB} \ket{\Phi^0_\rmA} + O_{\rmB\rmB} \ket{\Phi^0_\rmB} \quad.
\end{align}
\end{subequations}
Clearly, both $\{\hat{P}^0\ket{\Psi_0},\hat{P}^0\ket{\Psi_1}\}$ and $\{\hat{P}^0\ket{\Phi_\rmA},\hat{P}^0\ket{\Phi_\rmB}\}$ belong to the model subspace, but none of them are orthonormal subsets within it. 
However, they are expected to define two linearly independent families that span it. 
It must be understood that, as state vectors, they incarnate $\bold{\breve{C}}$ and $\bold{\breve{O}}$ viewed as their respective CI-coefficient matrix representations. 
This is one way to start effective Hamiltonian theory.

\subsection{From the model to the target}

In fact, we may proceed the other way around and rather invoke the 
projector onto the target subspace,
\begin{align}
\hat{P}
&= \ket{\Psi_0} \bra{\Psi_0} +\ket{\Psi_1} \bra{\Psi_1} \\ \nonumber
&= \ket{\Phi_\rmA} \bra{\Phi_\rmA} +\ket{\Phi_\rmB} \bra{\Phi_\rmB} \quad,
\end{align}
such that
\begin{subequations}
\begin{align}
& \hat{P} \ket{\Phi^0_\rmA} = C^*_{\rmA0} \ket{\Psi_0} + C^*_{\rmA1} \ket{\Psi_1} \quad, \\
& \hat{P} \ket{\Phi^0_\rmB} = C^*_{\rmB0} \ket{\Psi_0} + C^*_{\rmB1} \ket{\Psi_1} \quad,
\end{align}
\end{subequations}
or
\begin{subequations}
\label{eq:diatopcsf}
\begin{align}
& \hat{P} \ket{\Phi^0_\rmA} = O^*_{\rmA\rmA} \ket{\Phi_\rmA} + O^*_{\rmA\rmB} \ket{\Phi_\rmB} \quad, \\
& \hat{P} \ket{\Phi^0_\rmB} = O^*_{\rmB\rmA} \ket{\Phi_\rmA} + O^*_{\rmB\rmB} \ket{\Phi_\rmB} \quad.
\end{align}
\end{subequations}
Again, $\{\hat{P}\ket{\Phi^0_\rmA},\hat{P}\ket{\Phi^0_\rmB}\}$ is not an orthonormal subset within the target subspace. 
However, it is expected to define a linearly independent family within it, represented in matrix form either by $\bold{\breve{C}}^\dag$ or $\bold{\breve{O}}^\dag = \bold{\breve{T}} \bold{\breve{C}}^\dag$ (where $\bold{\breve{T}}^{-1} = \bold{\breve{T}}^\dag$).
This is another way to start effective Hamiltonian theory, which provides a useful formal frame and has been the subject of a vast literature in this context. 
In what follows, we shall stick to a more pedestrian approach based on relations fulfilled by full matrices and submatrices. 

\subsection{Invariant but nontrivial metric connection between both subspaces}

Now, let us make things more concrete.
The intermediate target set $\{\ket{\Phi_\rmA},\ket{\Phi_\rmB}\}$ obtained from the block-diagonalization procedure is 
orthonormal and is our focus here. We expect it to be quite close to the projected model set $\{\hat{P}\ket{\Phi^0_\rmA},\hat{P}\ket{\Phi^0_\rmB}\}$. 

Hence, we may assume the existence of a nonunitary, yet nonsingular, reciprocal-Jacobian-type matrix (model-to-target), $\bold{\breve{J}}$ such that 
\begin{subequations}
\label{eq:pcsftodia}
\begin{align}
& \ket{\Phi_\rmA} = J_{\rmA\rmA} \hat{P}\ket{\Phi^0_\rmA} + J_{\rmB\rmA} \hat{P}\ket{\Phi^0_\rmB} \quad, \\
& \ket{\Phi_\rmB} = J_{\rmA\rmB} \hat{P}\ket{\Phi^0_\rmA} + J_{\rmB\rmB} \hat{P}\ket{\Phi^0_\rmB} \quad.
\end{align}
\end{subequations}
Eq.~(\ref{eq:pcsftodia}) is reciprocal to Eq.~(\ref{eq:diatopcsf}), so that $\bold{\breve{J}}$ clearly satisfies
\begin{equation}
\bold{\breve{J}}^{-1} = \bold{\breve{O}}^\dag = \bold{\breve{T}} \bold{\breve{C}}^\dag \quad.
\end{equation}

Then, let us define the Gram (overlap) metric matrix of $\{\hat{P}\ket{\Phi^0_\rmA},\hat{P}\ket{\Phi^0_\rmB}\}$, 
\begin{equation}
\bold{\breve{S}} = 
\begin{pmatrix}
\mel{\Phi^0_\rmA}{\hat{P}}{\Phi^0_\rmA} & \mel{\Phi^0_\rmA}{\hat{P}}{\Phi^0_\rmB} \\
\mel{\Phi^0_\rmB}{\hat{P}}{\Phi^0_\rmA} & \mel{\Phi^0_\rmB}{\hat{P}}{\Phi^0_\rmB} 
\end{pmatrix} \quad,
\end{equation}
where we used the idempotency of the projector, $\hat{P}^2=\hat{P}$. 
Being the matrix representation of the projector, it can also be viewed as the ensemble density matrix of the target subspace with respect to the model subspace. 
It is Hermitian, expected to be positive definite, and satisfies
\begin{equation}
\bold{\breve{S}} = (\bold{\breve{J}}\bold{\breve{J}}^\dag)^{-1} = \bold{\breve{O}}\bold{\breve{O}}^\dag = \bold{\breve{C}}\bold{\breve{C}}^\dag \quad.
\end{equation}
Clearly, the $\bold{\breve{S}}$-matrix is a metric invariant, which only depends on the nature of -- and connection between -- the model and the target subspaces, and not on the particular choice of a specific subset of state vectors.
From there, we obtain the generalized orthonormality and partial closure relations within the target subspace as follows,
\begin{subequations}
\begin{align}
& \bm1 = \bold{\breve{J}}^\dag \bold{\breve{S}} \bold{\breve{J}} \quad, \\
& \hat{P} = \sum_{J\in\{\rmA,\rmB\}}\sum_{K\in\{\rmA,\rmB\}} \hat{P}\ket{\Phi^0_J}
[\bold{\breve{S}}^{-1}]_{JK}
\bra{\Phi^0_K}\hat{P} \quad. 
\end{align}
\end{subequations}

\subsection{Löwdin Hermitian orthonormalization}

Because $\bold{\breve{S}}$ is Hermitian and positive definite, there exist both a unitary matrix $\bold{\breve{U}}$ and a complex-diagonal matrix $\bold{\breve{\Sigma}}$ such that (unitary diagonalization)
\begin{equation}
\bold{\breve{S}} = \bold{\breve{U}}  \bold{\breve{\Sigma}}^2 \bold{\breve{U}}^\dag \quad, \quad \bm1 = \bold{\breve{U}}\bold{\breve{U}}^\dag \quad,
\end{equation}
whereby the $\bold{\breve{\Sigma}}^2$-matrix is real-diagonal and positive definite. 
For future convenience (related to SVD conventions) we shall assume that the positive eigenvalues within $\bold{\breve{\Sigma}}^2$ are sorted in nonincreasing order, which fully determines it.
Further, we can assume the $\bold{\breve{\Sigma}}$-matrix to be uniquely determined and chosen as positive definite too, such that $\bold{\breve{\Sigma}}=|\bold{\breve{\Sigma}}|=(\bold{\breve{\Sigma}}^2)^{1/2}$ [the half-exponent notation denotes the unique positive semidefinite principal square root of a positive semidefinite matrix].
However, the definition of $\bold{\breve{U}}$ is not unique and is up to a diagonal unitary matrix of phase factors that commutes with $\bold{\breve{\Sigma}}^2$.

From the above, the $\bold{\breve{J}}$-matrix is not unique either: it is defined up to multiplications with an infinite/continuous Lie group of undetermined unitary matrices.
The Löwdin Hermitian (symmetric if real)  orthonormalization  procedure proposes a well-determined, and well-known, least-transformed optimal solution -- denoted from now on with a `five-star' subscript index -- under the following Hermitian and positive definite form, 
\begin{equation}
\bold{\breve{J}}_\star = \bold{\breve{S}}^{-1/2} = \bold{\breve{U}}  \bold{\breve{\Sigma}}^{-1} \bold{\breve{U}}^\dag \quad,
\end{equation}
\emph{i.e.}, 
\begin{equation}
\bold{\breve{J}}_\star  =  (\bold{\breve{J}}_\star\bold{\breve{J}}_\star^\dag)^{1/2} = \bold{\breve{J}}_\star^\dag \quad.
\end{equation}

This is equivalent for the optimal $\bold{\breve{O}}_\star$ to satisfy 
\begin{equation}
\bm1 = \bold{\breve{O}}_\star^\dag (\bold{\breve{O}}_\star\bold{\breve{O}}^\dag_\star)^{-1/2}
= \bold{\breve{O}}_\star^\dag \bold{\breve{S}}^{-1/2} \quad,
\end{equation}
so that
\begin{equation}
\label{eq:propertyo}
\bold{\breve{O}}_\star=\bold{\breve{S}}^{1/2}=(\bold{\breve{O}}_\star\bold{\breve{O}}^\dag_\star)^{1/2}=\bold{\breve{O}}^{\dag}_\star \quad.
\end{equation}
Clearly, $\bold{\breve{O}}_\star$ is an invariant. We shall see later on that it determines optimally quasi-diabatic states (``as diabatic as possible'').

Then, from $\bold{\breve{T}}^\dag_\star = \bold{\breve{C}}^\dag \bold{\breve{O}}^{\dag-1}_\star=\bold{\breve{C}}^\dag \bold{\breve{J}}_\star$, we get
\begin{equation}
\bold{\breve{T}}^\dag_\star = \bold{\breve{C}}^\dag (\bold{\breve{C}}\bold{\breve{C}}^\dag)^{-1/2} \quad,
\end{equation}
\emph{i.e.}, $\bold{\breve{T}}_\star$ is constructed as a unitary matrix that is as close as possible to $\bold{\breve{C}}$, which occurs to be exactly the proposition of Werner \emph{et al.} for determining the optimal ADT matrix from the knowledge of $\bold{\breve{C}}$ \cite{wer88:3139,sim99:4523}.

\subsection{Unitary Procrustes analysis}

We shall now explore an SVD reformulation of the previous problem. 
From the above, there further exists some unitary matrix $\bold{\breve{V}}$ such that
\begin{equation}
\bold{\breve{S}} = (\bold{\breve{U}}  \bold{\breve{\Sigma}})(\bold{\breve{\Sigma}}\bold{\breve{U}}^\dag) = (\bold{\breve{U}}  \bold{\breve{\Sigma}}\bold{\breve{V}}^\dag)(\bold{\breve{V}}
\bold{\breve{\Sigma}}\bold{\breve{U}}^\dag) = \bold{\breve{C}}\bold{\breve{C}}^\dag \quad,
\end{equation}
where (square SVD)
\begin{equation}
\bold{\breve{C}} = \bold{\breve{U}}  \bold{\breve{\Sigma}}\bold{\breve{V}}^\dag \quad.
\end{equation}
The $\bold{\breve{\Sigma}}$-matrix is not the identity but is expected to be close to it (according to the quality of the model \emph{vs.} target) and the singular values are necessarily to be found between one and zero (degradation from coincidence to full mismatch). 
As shown below, there are two equivalent ways of formulating our objective, either as a maximization or a minimization problem. 

First, we are searching for a unitary matrix $\bold{\breve{T}}_\star$ that satisfies
\begin{equation}
\bold{\breve{T}}_\star := \arg \max_{\bold{\breve{T}}=\bold{\breve{T}}^{\dag-1}} \mathfrak{R}\trace{(\bold{\breve{C}}^\dag\bold{\breve{T}})} \quad,
\end{equation}
expressed in terms of the maximization of the real part of a Frobenius matrix inner product, $\bold{\breve{C}}:\bold{\breve{T}} = \trace{(\bold{\breve{C}}^\dag\bold{\breve{T}})}$.
From the square SVD and the cyclic-permutation invariance property of the trace, we have
\begin{equation}
\trace{(\bold{\breve{C}}^\dag\bold{\breve{T}})} = \trace{(\bold{\breve{V}}\bold{\breve{\Sigma}}\bold{\breve{U}}^\dag\bold{\breve{T}})} = \trace{(\bold{\breve{\Sigma}}\bold{\breve{U}}^\dag\bold{\breve{T}}\bold{\breve{V}})} \quad,
\end{equation}
hence
\begin{equation}
\mathfrak{R}\trace{(\bold{\breve{C}}^\dag\bold{\breve{T}})} \leq \trace{(\bold{\breve{\Sigma}})} \quad,
\end{equation}
because $\bold{\breve{U}}^\dag\bold{\breve{T}}\bold{\breve{V}}$ is unitary. 

An obvious optimum, $\trace{(\bold{\breve{C}}^\dag\bold{\breve{T}})} = \trace{(\bold{\breve{\Sigma}})}$, is reached when 
\begin{equation}
\bold{\breve{T}} = \bold{\breve{U}} \bold{\breve{V}}^\dag =: \bold{\breve{T}}_\star \quad.
\end{equation}
It must be noted that this is an alternative, less usual, yet well-known, formulation of the Löwdin Hermitian orthonormalization problem (see above). 
Indeed,
\begin{align}
\label{eq:adtcondi}
\bold{\breve{T}}_\star
&= \bold{\breve{U}} \bold{\breve{V}}^\dag \\ \nonumber
&= \bold{\breve{U}}  \bold{\breve{\Sigma}}^{-1} \bold{\breve{U}}^\dag \bold{\breve{U}}  \bold{\breve{\Sigma}}\bold{\breve{V}}^\dag \\ \nonumber
&= (\bold{\breve{C}}\bold{\breve{C}}^\dag)^{-1/2}\bold{\breve{C}} \quad.
\end{align}

Second, the same objective can be reformulated as a unitary Procrustes problem (finding the nearest unitary matrix to some nonunitary matrix). 
Here, we are looking for a unitary matrix $\bold{\breve{T}}_\star$ such that
\begin{equation}
\bold{\breve{T}}_\star := \arg \min_{\bold{\breve{T}} = \bold{\breve{T}}^{\dag-1}} \norm{\bold{\breve{C}}-\bold{\breve{T}}} \quad,
\end{equation}
where the matrix Frobenius norm of any matrix $\bold{R}$ is defined as 
\begin{equation}
\norm{\bold{R}}=\sqrt{\bold{R}:\bold{R}}=\sqrt{\trace{(\bold{R}^\dag \bold{R}})} \quad.
\end{equation}
Since
\begin{align}
\norm{\bold{\breve{C}}-\bold{\breve{T}}}^2 &= \norm{\bold{\breve{C}}}^2 + \norm{\bold{\breve{T}}}^2 - 2\mathfrak{R}\trace{(\bold{\breve{C}}^\dag\bold{\breve{T}})} \\ \nonumber
&= \trace{(\bold{\breve{\Sigma}}^2)} + \trace{(\bm1)} - 2\mathfrak{R}\trace{(\bold{\breve{C}}^\dag\bold{\breve{T}})} \\ \nonumber
&\geq \trace{(\bold{\breve{\Sigma}}^2)} + \trace{(\bm1)} - 2\trace{(\bold{\breve{\Sigma}})} \quad,
\end{align}
we observe that maximizing $\mathfrak{R}\trace{(\bold{\breve{C}}^\dag\bold{\breve{T}})}$ is equivalent to minimizing $\norm{\bold{\breve{C}}-\bold{\breve{T}}}$.
We obviously obtain the same optimum as above, denoted $\bold{\breve{T}}_\star$, such that the minimal (least-transformed) matrix distance is
\begin{equation}
\norm{\bold{\breve{C}}-\bold{\breve{T}}_\star} = \norm{\bold{\breve{\Sigma}}-\bm1} \quad.
\end{equation}
Hence, the standard deviation of the singular values with respect to one is an objective measure of the distance from the matrix $\bold{\breve{C}}$ to its nearest unitary matrix $\bold{\breve{T}}_\star$. 
This magnitude is incompressible and can be viewed as the intrinsic shortest-pathway distance between the model and the target subspaces. 

\subsection{Reverse polar decomposition}

The previous analysis was focused on $\bold{\breve{C}}$ for constructing $\bold{\breve{T}}$. For us, $\bold{\breve{T}}$ is not an objective but somewhat of a by-product obtained from the knowledge of a given $\bold{\breve{O}}$ for producing $\bold{\breve{C}}$.
Thus, we are rather interested in the properties of $\bold{\breve{O}}$. 
Much as the former, the latter admits a square SVD,
\begin{equation}
\bold{\breve{O}} = \bold{\breve{C}}\bold{\breve{T}}^\dag = \bold{\breve{U}}  \bold{\breve{\Sigma}}\bold{\breve{W}}^\dag \quad,
\end{equation}
where the right unitary matrix thus satisfies
\begin{equation}
\bold{\breve{W}} = \bold{\breve{T}}\bold{\breve{V}} \quad.
\end{equation}
Hence, both $\bold{\breve{O}}$ and $\bold{\breve{C}}$ share the same singular values by construction, irrespective of anything else. 

We can introduce a unitary matrix termed $\bold{\breve{B}}$, such that 
\begin{equation}
\bold{\breve{O}} = \bold{\breve{O}}_\star\bold{\breve{B}} \quad,
\end{equation}
where, again, $\bold{\breve{O}}_\star$ is a Hermitian matrix.
This is known as a reverse (or left) polar decomposition and is such that
\begin{equation}
\bold{\breve{O}} = (\bold{\breve{U}}  \bold{\breve{\Sigma}}\bold{\breve{U}}^\dag)(\bold{\breve{U}}\bold{\breve{W}}^\dag) \quad,
\end{equation}
where
\begin{subequations}
\begin{align}
&\bold{\breve{O}}_\star = \bold{\breve{U}}  \bold{\breve{\Sigma}}\bold{\breve{U}}^\dag = \bold{\breve{S}}^{1/2} \quad, \\
&\bold{\breve{B}} = \bold{\breve{U}}\bold{\breve{W}}^\dag  = \bold{\breve{U}} \bold{\breve{V}}^\dag \bold{\breve{T}}^\dag \quad.
\end{align}
\end{subequations}
As already alluded to, an ``optimally diabatic'' $\bold{\breve{O}}$-matrix should be such that the $\bold{\breve{B}}$-matrix boils down to the identity matrix. This is equivalent to ensuring that $\bold{\breve{T}}=\bold{\breve{U}} \bold{\breve{V}}^\dag$ (see above) or to having $\bold{\breve{W}}=\bold{\breve{U}}$. 

Our central point here is that we expect the variational SA-VQE algorithm to have minimally travelled along the shortest-distance pathway within the $\bmtheta$-parameter landscape (least-action principle) until it found the block-diagonalization solution given by $\bold{\breve{O}}$, expected to be as close as possible to the optimal $\bold{\breve{O}}_\star$ according to the proximity of $\bold{\breve{B}}$ with $\bm1$.

\subsection{From submatrices to full partitioning}

Our discussion so far has been focussed on the submatrices restricted to the model and target subspaces.
In order to make the block-diagonalizing partition more concrete, we can express the full, unitary, $\bold{O}$-matrix as follows,
\begin{equation}
\bold{O} = 
\begin{pmatrix}
\bold{\breve{O}} & \bold{Y} \\
\bold{X} & \bold{\bar{Z}}
\end{pmatrix} \quad.
\end{equation}
While $\bold{Y}$ and $\bold{\bar{Z}}$ are not to be determined explicitly, we can examine the properties of the target-model-mixing $\bold{X}$-matrix. 
This is helped by the fact that we are considering a bipartition (single target subspace and its complement). 
Thanks to this, we can rely on the conservation of $\bold{\breve{O}}^\dag\bold{\breve{O}} + \bold{X}^\dag\bold{X} = \bm1$.
In other words, maximizing the closeness of $\bold{\breve{O}}$ with the identity automatically ensures the closeness of $\bold{X}$ with zero. 
This can be expressed more concretely as
\begin{equation}
\trace{(
\begin{pmatrix}
\bold{\breve{O}}^\dag & \bold{\breve{X}}^\dag
\end{pmatrix}
\begin{pmatrix}
\bm1 \\
\bm0
\end{pmatrix}
)} = \trace{(\bold{\breve{O}}^\dag\bm1)} \quad,
\end{equation}
whereby we want
\begin{equation}
\label{eq:easycond}
\begin{pmatrix}
\bold{\breve{O}}_\star \\
\bold{X}_\star
\end{pmatrix}
\approx
\begin{pmatrix}
\bm1 \\
\bm0
\end{pmatrix} \quad,
\end{equation}
which is consistent with the proposition of Werner \emph{et al.} \cite{wer88:3139,sim99:4523}. 
We shall come back to this later on when addressing the definition of the diabatic reference. 

The approach proposed by Cederbaum \emph{et al.} \cite{pac88:7367,ced89:2427} is somewhat equivalent but more involved\martinfix{,}{} since it deals with the exact block-partitioning of the full $\bold{O}$-matrix from the onset.
The complete derivation is quite thorough, yet nicely exposed in the aforementioned references. 
Let us try to summarize their most important results, with some assumed paraphrasing, but revisited with our notations and complementary derivations. 

Using the notations defined in Appendix~\ref{app:NAC}, 
the four blocks of the unitary $\bold{C}$-matrix (CI ``partial eigenvectors'': eigenvectors only in its first two columns; unitary complement in the others) satisfy 
\begin{equation}
\bold{C} = \bold{O}\bold{T} =
\begin{pmatrix}
\bold{\breve{O}} & \bold{Y} \\
\bold{X} & \bold{\bar{Z}}
\end{pmatrix} 
\begin{pmatrix}
\bold{\breve{T}} & \bm0 \\
\bm0 & \bm1
\end{pmatrix} =
\begin{pmatrix}
\bold{\breve{O}}\bold{\breve{T}} & \bold{Y} \\
\bold{X}\bold{\breve{T}} & \bold{\bar{Z}}
\end{pmatrix} \quad,
\end{equation}
where $\bold{\breve{O}}\bold{\breve{T}}=\bold{\breve{C}}$. 
From the unitarity of $\bold{C}$ or $\bold{O}$, we get, in particular,
\begin{subequations}
\begin{align}
&\bold{\breve{T}}^\dag\bold{\breve{O}}^\dag\bold{Y} + \bold{\breve{T}}^\dag\bold{X}^\dag\bold{\bar{Z}} = \bm0 \quad, \\
&\bold{\breve{O}}^\dag\bold{Y} + \bold{X}^\dag\bold{\bar{Z}} = \bm0 \quad.
\end{align}
\end{subequations}
Hence, defining the following rectangular matrix,
\begin{equation}
\label{eq:rectangularl}
\boldsymbol{\lambda} = \bold{X}\bold{\breve{O}}^{-1} = \bold{X}\bold{\breve{T}}\bold{\breve{C}}^{-1} = -(\bold{Y}\bold{\bar{Z}}^{-1})^\dag \quad,
\end{equation}
we can recast $\bold{C}$ and $\bold{O}$ as
\begin{equation}
\bold{C} =
\begin{pmatrix}
\bold{\breve{C}} & -\boldsymbol{\lambda}^\dag\bold{\bar{Z}} \\
\boldsymbol{\lambda}\bold{\breve{C}} & \bold{\bar{Z}}
\end{pmatrix} \quad, \quad
\bold{O} =
\begin{pmatrix}
\bold{\breve{O}} & -\boldsymbol{\lambda}^\dag\bold{\bar{Z}} \\
\boldsymbol{\lambda}\bold{\breve{O}} & \bold{\bar{Z}}
\end{pmatrix} \quad.
\end{equation}
As such, the $\boldsymbol{\lambda}$-matrix can be viewed as a relative measure of the extent of the off-diagonal blocks with respect to on-diagonal blocks.
It is closely related to the ``small parameter'' in perturbation theory. 
It must be stressed that $\boldsymbol{\lambda}$ is an invariant matrix with respect to block-diagonalization. 
It is thus an incompressible measure of the quality of the model \emph{vs.} the target.

From there, we can decompose the previous matrices as
\begin{equation}
\label{eq:defmmat}
\bold{C} = \bold{M}
\begin{pmatrix}
\bold{\breve{C}} & \bm0 \\
\bm0 & \bold{\bar{Z}}
\end{pmatrix} \quad, \quad
\bold{O} = \bold{M}
\begin{pmatrix}
\bold{\breve{O}} & \bm0 \\
\bm0 & \bold{\bar{Z}}
\end{pmatrix} \quad,
\end{equation}
\emph{i.e.},
\begin{subequations}
\begin{align}
&\bold{C} = 
\begin{pmatrix}
\bold{\breve{C}} & \bm0 \\
\bm0 & \bold{\bar{Z}}
\end{pmatrix} 
+ \begin{pmatrix}
\bm0 & -\boldsymbol{\lambda}^\dag \bold{\bar{Z}} \\
\boldsymbol{\lambda}\bold{\breve{C}} & \bm0
\end{pmatrix}
\quad, \\
&\bold{O} = 
\begin{pmatrix}
\bold{\breve{O}} & \bm0 \\
\bm0 & \bold{\bar{Z}}
\end{pmatrix}
+ \begin{pmatrix}
\bm0 & -\boldsymbol{\lambda}^\dag \bold{\bar{Z}} \\
\boldsymbol{\lambda}\bold{\breve{O}} & \bm0
\end{pmatrix} \quad,
\end{align}
\end{subequations}
where 
\begin{equation}
\bold{M} =
\begin{pmatrix}
\bm1 & -\boldsymbol{\lambda}^\dag \\
\boldsymbol{\lambda} & \bm1
\end{pmatrix}
=\bm1 + \bold{\Lambda}
\underset{\bold{\Lambda} \rightarrow \bm0}{\sim} e^{\bold{\Lambda}} \quad,
\end{equation}
with
\begin{equation}
\bold{\Lambda} =
\begin{pmatrix}
\bm0 & -\boldsymbol{\lambda}^\dag \\
\boldsymbol{\lambda} & \bm0
\end{pmatrix}
\quad.
\end{equation}
The $\bold{M}$-matrix somewhat represents an intermediate-normalized first-order expansion in terms of the skew-Hermitian $\bold{\Lambda}$-matrix. 
As for $\boldsymbol{\lambda}$, let us stress that both $\bold{M}$ and $\bold{\Lambda}$ are invariant matrices.

The adjoint of $\bold{M}$ reads
\begin{equation}
\bold{M}^\dag =
\begin{pmatrix}
\bm1 & \boldsymbol{\lambda}^\dag \\
-\boldsymbol{\lambda} & \bm1
\end{pmatrix}
= 2\bm1-\bold{M} =\bm1 - 
\bold{\Lambda} 
\underset{\bold{\Lambda} \rightarrow \bm0}{\sim} e^{-\bold{\Lambda}}\quad,
\end{equation}
and obvioulsy commutes with $\bold{M}$. We thus have
\begin{align}
\bold{M}^\dag\bold{M} = \bold{M}\bold{M}^\dag
&= 2\bold{M} - \bold{M}^2 \\ \nonumber
&= \bm1 - \bold{\Lambda}^2
= \begin{pmatrix}
\bm1+\boldsymbol{\lambda}^\dag\boldsymbol{\lambda} & \bm0 \\
\bm0 & \bm1+\boldsymbol{\lambda}\boldsymbol{\lambda}^\dag
\end{pmatrix} \quad.
\end{align}

While $\bold{M}$ was used to produce the full CI-coefficient matrices from their block-diagonal restrictions [see Eq.~(\ref{eq:defmmat})], its inverse realizes the reverse mapping,
\begin{equation}
\begin{pmatrix}
\bold{\breve{C}} & \bm0 \\
\bm0 & \bold{\bar{Z}}
\end{pmatrix}
= \bold{M}^{-1}\bold{C} \quad, \quad \begin{pmatrix}
\bold{\breve{O}} & \bm0 \\
\bm0 & \bold{\bar{Z}}
\end{pmatrix}
= \bold{M}^{-1}\bold{O} \quad.
\end{equation}
It satisfies 
\begin{align}
\bold{M}^{-1}
&= (\bm1 + 
\bold{\Lambda})^{-1} \\ \nonumber
&= (\bm1 - 
\bold{\Lambda})(\bm1 - \bold{\Lambda}^2)^{-1} \\ \nonumber
&= (\bm1 - \bold{\Lambda}^2)^{-1}
(\bm1 - 
\bold{\Lambda}) \\ \nonumber
&= (\bm1 - \bold{\Lambda}^2)^{-1/2}
(\bm1 - 
\bold{\Lambda})(\bm1 - \bold{\Lambda}^2)^{-1/2} 
\underset{\bold{\Lambda} \rightarrow \bm0}{\sim} e^{-\bold{\Lambda}} \quad.
\end{align}
The following commutative product
\begin{equation}
\bold{M}^\dag\bold{M}^{-1} = (\bm1 - 
\bold{\Lambda})(\bm1 + 
\bold{\Lambda})^{-1} 
\underset{\bold{\Lambda} \rightarrow \bm0}{\sim} e^{-2\bold{\Lambda}} \quad,
\end{equation}
is a unitary matrix known as the Cayley transform of $\bold{\Lambda}$. It provides an approximation of the exponential, exact to second order in the exponent, also known as the Crank-Nicolson scheme in the context of propagators. Further,
\begin{equation}
(\bold{M}^\dag\bold{M}^{-1})^{-1/2} = \bold{M}(\bold{M}^\dag\bold{M})^{-1/2}
\underset{\bold{\Lambda} \rightarrow \bm0}{\sim} e^{\bold{\Lambda}} \quad,
\end{equation}
is a unitary matrix that approximates $\bold{M}$ to first order in $\bold{\Lambda}$ and thus provides an optimal solution to $\bold{O}$ for $\bold{\breve{O}}$ and $\bold{\bar{Z}}$ to be as close as possible to identity matrices.
It is the renormalized form of $\bold{M}$ within the context of QDPT.

Hence, the least-transformation principle from model to target is stated for the full unitary matrix as follows,
\begin{equation}
\label{eq:leastaction}
\bold{O}_\star := \arg \min_{\bold{O} = \bold{O}^{\dag-1}} \norm{\bold{O}-\bm1} \quad.
\end{equation}
An optimal solution to Eq.~(\ref{eq:leastaction}) (least-transformed block-diagonalization) is thus given by
\begin{equation}
\bold{O}_\star=\bold{M}(\bold{M}^\dag\bold{M})^{-1/2} \quad,
\end{equation}
which agrees with our previous handwaving justification based on the Cayley transform. 
We shall show below that this also complies with what has been derived in the previous subsection for the target submatrices. 

As already stated, the $\bold{M}$-matrix is the intermediate-normalized contribution to $\bold{O}_\star$-matrix, while the $(\bold{M}^\dag\bold{M})^{-1/2}$-matrix ensures its renormalization.
As a result, we get the two terms of $\bold{O}_\star$ as
\begin{subequations}
\label{eq:formofowrtl}
\begin{align}
&\begin{pmatrix}
\bold{\breve{O}}_\star & \bm0 \\
\bm0 & \bold{\bar{Z}}
\end{pmatrix}
= (\bold{M}^\dag\bold{M})^{-1/2}
= (\bm1 - \bold{\Lambda}^2)^{-1/2} \quad, \\
&\begin{pmatrix}
\bm0 & \bold{Y} \\
\bold{X}_\star & \bm0
\end{pmatrix}
= \bold{\Lambda} \begin{pmatrix}
\bold{\breve{O}}_\star & \bm0 \\
\bm0 & \bold{\bar{Z}}
\end{pmatrix}
= \bold{\Lambda}(\bm1 - \bold{\Lambda}^2)^{-1/2} \quad.
\end{align}
\end{subequations}
Quite obviously, $\bold{O}_\star$ tends to $\bm1$ if $\bold{\Lambda}$ tends to $\bm0$ (to second order in $\bold{\Lambda}$ for the on-diagonal submatrices and to first order for the off-diagonal ones).
It must be noted that this provides a particular unitary parametrization for $\bold{O}_\star$ whereby its on-diagonal submatrices are assumed to be Hermitian, as shown in the previous subsection. 

Now, in agreement with our previous definition of the Gram matrix, we have the following Hermitian and positive definite matrix invariants,
\begin{subequations}
\label{eq:blockgram}
\begin{align}
&\bold{\breve{S}}=\bold{\breve{C}}\bold{\breve{C}}^\dag = \bold{\breve{O}}\bold{\breve{O}}^\dag = \bold{\breve{O}}_\star\bold{\breve{O}}^\dag_\star = (\bm1+\boldsymbol{\lambda}^\dag\boldsymbol{\lambda})^{-1} \quad, \\
&\bold{\bar{S}}=\bold{\bar{Z}}\bold{\bar{Z}}^\dag = (\bm1+\boldsymbol{\lambda}\boldsymbol{\lambda}^\dag)^{-1} \quad.
\end{align}
\end{subequations}
In particular, this holds whether $\bold{\breve{O}}$ is optimal or not.

Then, in the case of the optimal solution [see Eq.~(\ref{eq:formofowrtl})], this yields 
\begin{equation}
\begin{pmatrix}
\bold{\breve{O}}_\star \\
\bold{X}_\star
\end{pmatrix}
= \begin{pmatrix}
\bold{\breve{S}}^{1/2} \\
\boldsymbol{\lambda}\bold{\breve{S}}^{1/2}
\end{pmatrix} 
= \begin{pmatrix}
(\bm1+\boldsymbol{\lambda}^\dag\boldsymbol{\lambda})^{-1/2} \\
\boldsymbol{\lambda}(\bm1+\boldsymbol{\lambda}^\dag\boldsymbol{\lambda})^{-1/2} \\
\end{pmatrix} \quad,
\end{equation}
which is consistent with Eq.~(\ref{eq:easycond}) for a small $\boldsymbol{\lambda}$.
Once again, this holds if $\bold{\breve{O}}_\star$ is Hermitian.
As such, the simpler formulation of Werner \emph{et al.} and the more general one of Cederbaum \emph{et al.} are equivalent. Note that there remains some gauge freedom for $\bold{\bar{Z}}$ in practice, which is inconsequential for the present purpose (we can formally assume that it satisfies the optimal condition $\bold{\bar{Z}}=\bold{\bar{S}}^{1/2}=\bold{\bar{Z}}^\dag$).

\subsection{Relation with NACs and diabaticity}

Now, we should address the crucial question here: why is the least-transformation principle a guarantee of maximal diabaticity?
The justification is evident intuitively from examining its meaning in terms of maximal overlap or minimal residual between model and target state vectors.
Yet, a more detailed analysis of its consequences on NACs will help.
Once again, we shall follow the spirit of Cederbaum \emph{et al.}.
While they rely on the knowledge of some `crude adiabatic' basis set at a given reference geometry, the derivation is similar when trusting a CI ``good guess'' in terms of selected dominant CSFs, as shown by Werner \emph{et al.}, provided diabatic orbitals are available. This is the approach followed in the present work. 
In both cases, the essential point is that the model subspace is as diabatic as possible.

Most notations have been defined above, or in Appendix~\ref{app:NAC} and the main text. Let us recollect the most important CI-NAC relations in the full Hilbert space,
\begin{subequations}
\begin{align}
&\bold{F}^\mathrm{CI} = \bold{C}^\dag\bm\nabla\bold{C} 
= \bold{T}^\dag_\star\bold{f}^\mathrm{CI}_\star\bold{T}_\star + \bold{T}_\star^\dag\bm\nabla\bold{T}_\star \quad, \\
&\bold{f}^\mathrm{CI}_\star = \bold{O}_\star^\dag\bm\nabla\bold{O}_\star 
= \bold{T}_\star\bold{F}^\mathrm{CI}\bold{T}^\dag_\star + \bold{T}_\star\bm\nabla\bold{T}^\dag_\star \quad, \\
&\bold{f}^\mathrm{CI} = \bold{O}^\dag\bm\nabla\bold{O} 
= \bold{B}^\dag\bold{f}^\mathrm{CI}_\star\bold{B} + \bold{B}^\dag\bm\nabla\bold{B} \quad,
\end{align}
\end{subequations}
where the full-size unitary $\bold{B}$-matrix is constructed from $\bold{\breve{B}}$ much as $\bold{T}$ is constructed from $\bold{\breve{T}}$,
\begin{equation}
\bold{B} = 
\begin{pmatrix}
\bold{\breve{B}} & \bm0 \\
\bm0 & \bm1
\end{pmatrix} \quad, \quad
\bold{B}^\dag\bm\nabla\bold{B} = 
\begin{pmatrix}
\bold{\breve{B}}^\dag\bm\nabla\bold{\breve{B}} & \bm0 \\
\bm0 & \bm0
\end{pmatrix} \quad.
\end{equation}
We know that $\bold{f}^\mathrm{CI}_\star$ is an invariant that only depends on the relative definitions of the model and target subspaces. 
As such, it is an incompressible magnitude and thus qualifies as the minimal and nonremovable CI contribution to the NACs. 
Further, both $\bold{f}^\mathrm{CI}_\star$ and its unitary similarity transform $\bold{B}^\dag\bold{f}^\mathrm{CI}_\star\bold{B}$ have the same Frobenius norm and represent quantities of identical global magnitudes. 

On the other hand, if the extra term $\bold{B}^\dag\bm\nabla\bold{B}$ is nonzero, it adds up some extra contributions to the residual CI-NACs. 
It clearly is removable if one chooses $\bold{B}$ as the identity (see above for the particular choice that we made for the optimal solution), but other choices are equally acceptable as long as $\bold{B}$ does not depend on the nuclear coordinates at all ($\bm\nabla\bold{B} \equiv \bm0$). 
This corresponds to alternative matrix square roots that are not the principal ones.
Dealing with such a gauge freedom seems innocent but is actually a potential source of complication. 
In the present work, we obtained $\bold{B}$ as a matrix close to the identity and behaving in a uniform manner with respect to the nuclear coordinates (see main text), which is an important result in practice.

Now, let us examine more closely the expression of the nonremovable CI-NACs,
\begin{align}
\bold{f}^\mathrm{CI}_\star
&= (\bm1 - \bold{\Lambda}^2)^{-1/2}(\bm1 - \bold{\Lambda})\bm\nabla\bold{\Lambda}(\bm1 - \bold{\Lambda}^2)^{-1/2} \\ \nonumber
&+ (\bm1 - \bold{\Lambda}^2)^{1/2}\bm\nabla(\bm1 - \bold{\Lambda}^2)^{-1/2} \quad.
\end{align}
Such an expression should be reshuffled so as to get its block structure, but the presence of the gradient of the inverse of a matrix square root prevents any evident simplification. 
It perhaps is more illuminating to invoke a Taylor expansion, since $\bold{\Lambda}$ is supposed to be small. Indeed, we have 
\begin{align}
\bold{O}_\star
&=(\bm1 + \bold{\Lambda})(\bm1 - \bold{\Lambda}^2)^{-1/2} \\ \nonumber
&= \bm1 + \bold{\Lambda} + \frac{1}{2}\bold{\Lambda}^2 + \frac{1}{2}\bold{\Lambda}^3 + \mathcal{O}(\bold{\Lambda}^4) \quad,
\end{align}
\begin{align}
\bold{O}^\dag_\star
&=(\bm1 - \bold{\Lambda}^2)^{-1/2}(\bm1 - \bold{\Lambda}) \\ \nonumber
&= \bm1 - \bold{\Lambda} + \frac{1}{2}\bold{\Lambda}^2 - \frac{1}{2}\bold{\Lambda}^3 + \mathcal{O}(\bold{\Lambda}^4) \quad,
\end{align}
\begin{align}
\bm\nabla\bold{O}_\star
&= \bm\nabla\bold{\Lambda} + \frac{1}{2}(\bm\nabla\bold{\Lambda}\bold{\Lambda} + \bold{\Lambda}\bm\nabla\bold{\Lambda}) \\ \nonumber
&+ \frac{1}{2}(\bm\nabla\bold{\Lambda}\bold{\Lambda}^2+\bold{\Lambda}\bm\nabla\bold{\Lambda}\bold{\Lambda}+\bold{\Lambda}^2\bm\nabla\bold{\Lambda}) + \mathcal{O}(\bold{\Lambda}^3) \quad,
\end{align}
and
\begin{align}
\bold{f}^\mathrm{CI}_\star
&= \bold{O}_\star^\dag\bm\nabla\bold{O}_\star \\ \nonumber
&= \bm\nabla\bold{\Lambda} + \frac{1}{2}(\bm\nabla\bold{\Lambda}\bold{\Lambda} - \bold{\Lambda}\bm\nabla\bold{\Lambda}) \\ \nonumber
&+ \frac{1}{2} (\bm\nabla\bold{\Lambda}\bold{\Lambda}^2 + \bold{\Lambda}^2\bm\nabla\bold{\Lambda}) + \mathcal{O}(\bold{\Lambda}^3) \quad.
\end{align}
All terms are skew-Hermitian matrices. 

The zeroth-order inter-subspace term,
\begin{equation}
\bm\nabla\bold{\Lambda} =
\begin{pmatrix}
\bm0 & -\bm\nabla\boldsymbol{\lambda}^\dag \\
\bm\nabla\boldsymbol{\lambda} & \bm0
\end{pmatrix} \quad,
\end{equation}
is a measure of how valid the block-Born-Oppenheimer approximation is (between the target and its complement). Its magnitude reflects how much the model and target subspaces vary against each other to first order with respect to the nuclear coordinates. This could be measured via a Hellmann-Feynman formula showing that $\bm\nabla\boldsymbol{\lambda}$ is small if the typical energy gap between the target states and their complement is large.

The first-order intra-subspace term,
\begin{equation}
\frac{1}{2}(\bm\nabla\bold{\Lambda}\bold{\Lambda} - \bold{\Lambda}\bm\nabla\bold{\Lambda}) = \frac{1}{2}
\begin{pmatrix}
\boldsymbol{\lambda}^\dag\bm\nabla\boldsymbol{\lambda} - \bm\nabla\boldsymbol{\lambda}^\dag\boldsymbol{\lambda} & \bm0 \\
\bm0 & \boldsymbol{\lambda}\bm\nabla\boldsymbol{\lambda}^\dag - \bm\nabla\boldsymbol{\lambda}\boldsymbol{\lambda}^\dag
\end{pmatrix} \quad,
\end{equation}
remains small if both $\boldsymbol{\lambda}$ and its gradient are small. 
Hence, the nonremovable CI-NACs among the target states is expected to be negligible if $\boldsymbol{\lambda}$ takes small values and remains flat. 
The second-order term is an inter-subspace renormalization correction, while the next (third-order) term is an intra-subspace one, and so on.

Similar considerations hold for the second-order NACs that we have not addressed in the present work. We refer the reader to the work of Cederbaum \emph{et al.} \cite{pac88:7367,ced89:2427} for further details.

\subsection{A final word about the definition of the diabatic reference}

As shown above, even for the optimal quasi-diabatic representation, there are two sources of nonremovable NACs: $\bold{f}^\mathrm{CI}_\star = \bold{O}_\star^\dag\bm\nabla\bold{O}_\star$ (see above) and $\bold{f}^\mathrm{CSF}_\star = \bold{O}^\dag_\star\bold{f}^0\bold{O}_\star$. 
The former (CI-NAC) is essentially governed by $\bm\nabla\boldsymbol{\lambda}$, which means that the model does not have to be perfect (although small values of $\boldsymbol{\lambda}$ obviously help) but its quality has to remain as uniform as possible when varying the nuclear coordinates (small values of $\bm\nabla\boldsymbol{\lambda}$). 
Again, this reflects the extent of validity of the block-Born-Oppenheimer approximation. 
On the other hand, the latter (CSF-NAC) has essentially the same global magnitude as $\bold{f}^0$ and reflects the overall diabaticity of the CSFs and their underlying orbitals. 

In summary, there are essentially two types of strategies to reduce the quasi-diabatic NACs in the present context (least-transformed block-diagonalization). 
The first one, proposed by Cederbaum \emph{et al.} \cite{pac88:7367,ced89:2427} assumes a unique (frozen) model subspace, termed `crude adiabatic', determined at some reference geometry by a subset of local adiabatic eigenstates. 
As such, there is no longer any concept of CSF-NAC, and the whole problem fully relies on what we call CI-NACs. 
At the reference point, $\boldsymbol{\lambda}$ is zero by construction but $\bm\nabla\boldsymbol{\lambda}$ does not have to be so. 
There is no intra-subspace NAC within the model, but there is potentially some inter-subspace NAC with the complement (which can be neglected if the block-Born-Oppenheimer approximation holds).

Now, as soon as the nuclear coordinates vary, the frozen model will lose its quality with respect to the target and $\boldsymbol{\lambda}$ will increase (and quite fast). 
The only way to improve the situation is to increase the size of the frozen model subspace.
Such a strategy can be viewed as perturbative in spirit and is relevant for spectroscopic applications involving small-amplitude distortions around the reference geometry.
It also is of much interest if the latter is that of a conical intersection, so as to get a local quasi-diabatic representation that is consistent with the behavior of the adiabatic NACs around the degeneracy point where they are still large.
This aspect was explored quite extensively by Yarkony \emph{et al.} (see, \emph{e.g.}, Ref. \cite{wan19:9874} and many other references in this context), whereby they introduced a looser definition of the `crude adiabatic' representation (frozen CI coefficients but flexible orbitals and CSFs). 

The second strategy, proposed by Werner \emph{et al.} 
\cite{wer88:3139,sim99:4523}, which is closer to the present work, assumes a global and chemically-oriented choice of reference for the model, based on the nature of the dominant CSFs within the target eigenstates with an SA-MCSCF perspective, and which varies smoothly with the nuclear coordinates.
It relies on a trade-off between the nonremovable CSF-NACs and CI-NACs designed to temper the overall growth of the total NACs in the quasi-diabatic representation.

The CSF-NACs are not zero but minimized upon using diabatic orbitals.
The CI-NACs are assumed to remain small due to the adaptation of the model to the target with respect to the nuclear coordinates (which implies small variations of $\boldsymbol{\lambda}$). 
Hence, such a strategy can be viewed as variational in spirit and is perhaps more relevant for chemical applications involving large-amplitude distortions.

Somehow, and making an analogy with fluid mechanics, the first strategy is Eulerian in spirit (observing the NACs grow from a fixed point), while the second one is more Lagrangian (constraining the NACs to remain limited along the flow).

Finally, it must be understood that the approach that we exposed in the present work was not originally meant for producing a quasi-diabatic representation.
Because it relies on the definition of a ``good guess'' for starting a block-diagonalization procedure with SA-OO-VQE (which is an SA-MCSCF-type of method), we observed that it spontaneously enforces the prescriptions of the second strategy without any post-treatment, provided we used diabatic orbitals. 
As exemplified in the main text, the deviation from the optimal quasi-diabatic representation (measured by the $\bold{B}$-matrix) is negligible and brings a contribution to the NACs
of the same amplitude as the CSF contribution.

\newcommand{\Aa}[0]{Aa}

\end{document}